\documentclass{article}
\usepackage{amsmath,amsfonts,stackrel}
\usepackage{algorithmic}
\usepackage{algorithm}
\usepackage{array}
\usepackage{textcomp}
\usepackage{stfloats}
\usepackage{url}
\usepackage{verbatim}
\usepackage{graphicx}
\usepackage{caption}
\usepackage{subcaption}
\usepackage{cite}
\usepackage{color}
\usepackage{soul}
\usepackage{siunitx}
\usepackage{balance}
\usepackage{fullpage}
\usepackage{makecell}
\usepackage{tabu}
\usepackage{tablefootnote}
\usepackage{svg}

\DeclareMathOperator{\KL}{KL}
\newcommand{\rmd}{\mathrm{d}}

\begin{document}
\title{{AI-Guided Codesign Framework for Novel Material and Device Design applied to MTJ-based True Random Number Generators}}
\author{Karan P. Patel$^{1,*}$ \and Andrew Maicke$^{2,*}$  \and Jared Arzate$^2$ \and Jaesuk Kwon$^2$ \and  \and J. Darby Smith$^3$ \and James B. Aimone$^3$ \and Jean Anne C. Incorvia$^2$ \and Suma G. Cardwell$^3$ and Catherine D. Schuman$^{1,+}$}
\date{    $^1$University of Tennessee, Knoxville, TN, USA\\
    $^2$The University of Texas at Austin, Austin, TX, USA\\
        $^3$Sandia National Laboratories, Albuquerque, NM, USA\\
        $^*$These authors contributed equally.\\
        $^+$Corresponding author: cschuman@utk.edu\\[2ex]%
    }

\maketitle

\begin{abstract}
Novel devices and novel computing paradigms are key for energy-efficient, performant future computing systems. However, designing devices for new applications is often time-consuming and tedious. Here, we investigate the design and optimization of spin--orbit torque and spin transfer torque magnetic tunnel junction models as the probabilistic devices for true random number generation. We leverage reinforcement learning and evolutionary optimization to vary key device and material properties of the various device models for stochastic operation. Our AI-guided codesign methods generated different candidate devices capable of generating stochastic samples for a desired probability distribution, while also minimizing energy usage for the devices.
\end{abstract}

\section{Introduction}

\begin{figure}[t!]
    \centering
    \vskip -1cm 
    \includegraphics[width=\textwidth]{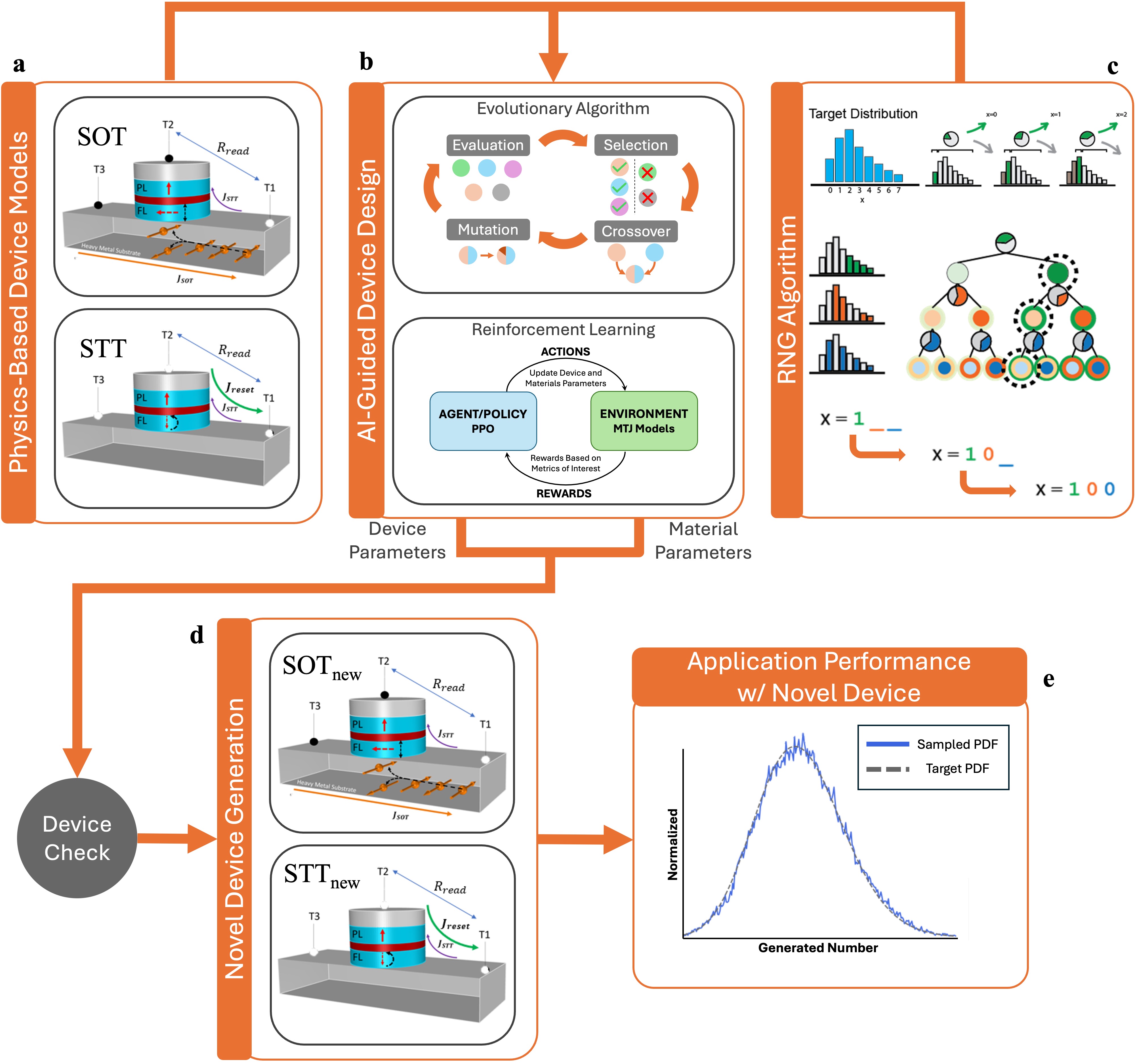} 
    \caption{Our AI-Guided Framework for Device Discovery and Optimization for a given application. Overview of the device model, AI-guided discovery and optimization strategy, and RNG algorithm workflow. Given a target distribution, the optimization approach (b) uses a device model (a) to simulate a true random bit according to the RNG algorithm (c). The optimization algorithm designs novel device configurations (d) that must pass device checks to be viable. The viable devices are used to produce the target distribution for a given application (e).}
    \label{fig:flowchart}
\end{figure}

Designing devices for novel applications is oftentimes a time rigorous and resource-constrained process that requires utilizing computationally intensive simulations, device fabrication, and testing of the physical components in the application-specific environment. At the same time, customizing device characteristics to a particular application can allow for significant performance improvements. Automated codesign strategies are becoming increasingly popular with advancements in the artificial intelligence (AI) field that provide useful machine learning algorithms and frameworks~\cite{zhang2020neuro, zhang2022,crowder2023ai,crowder2023deep}. Such codesign provides new opportunities to automatically customize devices for application-specific needs to maximize performance---whether that involves a particular capability, energy usage, latency, throughput, or even combinations of metrics. The operation of emerging devices, such as magnetic tunnel junctions (MTJs)~\cite{border2019integer,grimaldi2020single,kobayashi2021sigmoidal,ostwal2019spin}, can be simulated using physics-based models that capture key behaviors based on materials and device properties. By pairing these models with AI-guided codesign, we are able to effectively optimize the device parameters for application requirements and constraints~\cite{cardwell2022probabilistic,cardwell2024,schuman2023device}. 

AI-guided methods are increasingly being adopted in electronic design automation~(EDA) flows. Recently, reinforcement techniques have been used in EDA for multiple tasks including chip floor planning~\cite{mirhoseini2021graph}, architecture search~\cite{zhang2022}, gate sizing of VLSI~\cite{lu2021rl}, circuit optimization~\cite{budak2021dnn} and analog circuit design~\cite{settaluri2020autockt}. Evolutionary algorithm (EA) approaches, on the other hand, have been used for decades to design analog circuits~\cite{mattiussi2007analog} and can be creative in the design of novel solutions to a variety of problems~\cite{lehman2020surprising}. Both reinforcement learning (RL) and EA approaches are promising for optimization tasks, each offering unique pros and cons. In addition, recent work leverages generative AI (GAI)-based circuit characterization~\cite{benedetti2019generative} and optimization techniques~\cite{liang2023circuitops,chen2024dawn}. In related work, physics-informed neural networks (PINNs)~\cite{raissi2019physics}, originally designed for solving partial differential equations with informed loss functions, have been used to perform device design and optimization~\cite{lu2021physics,pan2023physics,knapp2022machine}. Codesign across devices, circuits, architectures, and applications for a full-stack solution is a challenge and an ongoing area of research. 

In previous work, we have shown initial results in leveraging RL for MTJ device codesign~\cite{cardwell2024} and EA for probabilistic circuit optimization using different MTJ devices and tunnel diode device~\cite{cardwell2022probabilistic}. This new work presents an intelligent, automated codesign framework for emerging devices.  In particular, we create a framework that is based on RL and EAs, which allows for multi-objective optimization of parameters of emerging devices for real-world applications.  We showcase this framework by providing a comparison of RL and EA approaches for device design and parameter optimization and a demonstration of device parameters for energy-efficient random number generation for gamma distributions for both spin--orbit torque (SOT) and spin transfer torque (STT) MTJ devices.  Though this framework is applied in the context of true random number generation using SOT and STT MTJ devices, it can be easily extended to other applications and other device types.

Ultimately, our methods produce the best candidate devices and materials properties for optimizing both performance in function and energy efficiency. Generally, we see that performance is improved but energy efficiency is slightly increased, compared to the default parameters used to represent standard CoFeB MTJs. The results also show that for the SOT MTJs, a larger range of material parameters can provide good performance, and material parameters for stronger perpendicular magnetic anisotropy (PMA) are favored. In contrast, for STT MTJs there is a narrower range of parameters to achieve the performance, and weaker PMA is favored.

This paper is structured as follows: in Section~\ref{TRNG} we provide details on our application with a background on RNGs and the distribution sampling scheme we employ, Section~\ref{Device_Models} introduces the two MTJ device types we will be designing for our application, Section~\ref{Optimizations} discusses our AI-guided approach using RL and EA  with Section~\ref{Results} presenting our AI-guided approaches results for device discovery. Section~\ref{Discussion} provides a discussion of the results and observations, and we finish with conclusions in Section~\ref{Conclusion}. Additional supplementary information is provided in Section~\ref{Supplementary_Info}.

\section{Application: True Random Number Generation for Non-Uniform Distributions}\label{TRNG}

To guide the development and discussion of our codesign framework, we focus on a single application --- true random number generation (TRNG), Fig.~\ref{fig:flowchart}c.  Despite this focus, we stress that our presented framework is general purpose for the AI-guided design and configuration of devices to meet application needs.  Here, our devices are both SOT and STT MTJ devices.  Our target function for these devices is to produce TRNG samples from a distribution of interest. 

Random number generation is a key component of many computational tasks, including scientific simulations, machine learning, and cryptography. In today's computing systems, random numbers are typically generated using pseudo-random number generators (PRNGs), which have several key limitations, including the quality of the random numbers generated (e.g., adherence to expected distributions or predictability) and periodicity of outcomes. Furthermore, many PRNGs are restricted to generating samples from a uniform distribution; if an application requires pseudo-random numbers from a different distribution, then additional computation is required to convert the sampled number to the appropriate distribution.

Recent studies into microelectronics and probabilistic computing have noted the need and potential for fast and efficient TRNG and noise sources~\cite{misra2022probabilistic,chowdhury2023full,aimone2024overcoming,daniels2023neural}. A step beyond random bits, as indicated in~\cite{misra2022probabilistic}, is to produce samples from specified distributions of interest, eliminating the need for expensive rejection sampling. This motivates our decision to focus solely on drawing from a specified distribution of interest.  While we elect to focus in on a single distribution, our prescribed framework can apply to any distribution that admits a tractable distribution. Additionally, since our devices are acting as biased coins, popular benchmarks to test the RNG quality, such as NIST, aren't applicable.

We choose to focus on a particular distribution in the gamma family.  The gamma distribution is a natural generalization of the exponential distribution.  It is applied in a variety of applications, including mathematical ecology (population and epidemic modeling), industrial systems engineering (queuing theory and related service time modeling), and finance (default modeling).  While the exponential distribution represents the waiting time of a single event arriving at a rate $\lambda$, the gamma distribution can be thought of as the waiting time for $n$ arrivals of events that individually arrive at a rate $\lambda$.  The two parameters of the gamma distribution are the shape $n$ and the rate $\lambda$. The exponential distribution arises as the special case when $n=1$. Both parameters can take on any positive value.

\begin{figure}[t!]
    \centering
    \vskip -3cm 
    \includegraphics[width=0.75\textwidth]{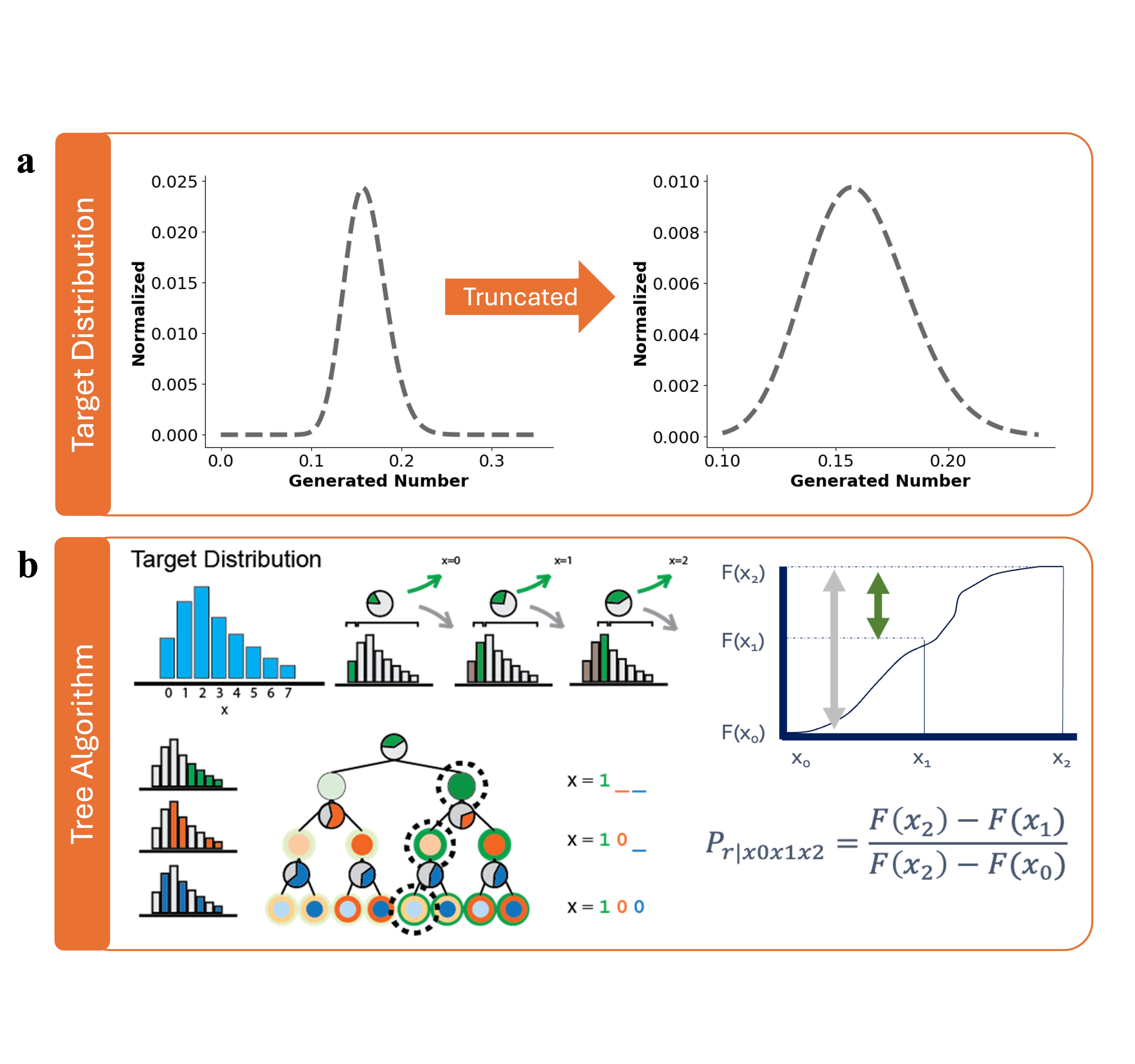} 
    \vskip -1cm
    \caption{Overview of tree algorithm. Truncated target distribution (a) is used in the tree algorithm (b) to optimize the MTJ devices to match the desired distribution.}
    \label{fig:tree_alg}
\end{figure}

For this effort, we selected $n=50.00$ and $\lambda=311.44$. This target distribution, shown in Figure~\ref{fig:tree_alg}a, provides features not present in a simple exponential. Namely, it provides an asymmetric shape tightly concentrated around a single value. These features and the entire codesign framework (presented in Sec.~\ref{Optimizations}) dramatically differentiate this work from existing literature applying vanilla RL to a vanilla exponential distribution~\cite{cardwell2024,cardwell2022probabilistic,maicke2023magnetic}. For explicit details on how we selected these values and an application to particle tracking, please see the Supplementary Material (Section~\ref{Supplementary_Info}).

In the context of our codesign workflow (Fig~\ref{fig:flowchart}), we will need an algorithm that transforms output from stochastic devices to samples from distributions of interest.  This is the RNG algorithm and it is the object (or function) of interest, informing the behavior of our devices and prescribing a need to precisely control their stochastic profile. In general, this choice of algorithm is a step in our framework for device design and can be arbitrary.

The RNG function we chose provides a binary-coded random number according to the desired probability density function (PDF) by utilizing output from tuned coinflips. Prescriptively, we begin with a PDF supported on some finite interval $[a,b]$. If the PDF is not finitely supported, we truncate it in a negligible way and renormalize. Our selected gamma distribution is not finitely supported. Hence, we will need to truncate the infinite support. We truncate our PDF to the interval $[0.10,0.24]$. This interval contains approximately $99.79\%$ of the mass of the distribution and is illustrated in Fig.~\ref{fig:tree_alg}a. Our selection for the bounds of the truncation was arbitrary, but our methodology would still be valid for any finite range. Note, with our RNG algorithm of choice (discussed in the following paragraph) there is no possibility of returning values outside the truncated range, even with noise imparted by the device.

Once we have an appropriate PDF, we construct a sampling decision tree by discretizing the PDF. For a $k$-bit number, we divide the chosen interval $[a,b]$ into $2^k$ equally-spaced bins.\footnote{Bins need not be equally spaced, but for ease of discussion, we assert that each bin has the same extent.} We create a binary flip tree encoding the outcomes of each of these bins from a sequence of coin tosses. The first coin toss will have a probability of tails equal to the integral of the PDF from $a$ to $(a+b)/2$. This partitions the probability of the first $2^{k-1}$ bins and the second $2^{k-1}$ bins. The second layer of the tree comprised two weighted coins. The first coin has a probability of tails equal to the integral of the PDF from $a$ to $(3a+b)/4$. The second coin has a probability of heads equal to the integral of the PDF from $(a+3b)/4$ to $b$. This process is continued until the final layer, where the integration is performed over the width of a single bin. In total, there will be $2^k-1$ weighted coins. In practice, the coin in the topmost layer is flipped; depending on the outcome of the first toss, a single coin in the second layer is selected to toss. This process continues.

In contrast to precomputing all $2^k-1$ coin weight values in the full probability tree---a process that grows exponentially with the desired bit precision---we instead modify the scheme and utilize an online process. We traverse through only those branches of the probability tree that are relevant for each sample by exploiting the cumulative distribution function (CDF). This scheme is suitable for any distribution for which the CDF can be defined. For our gamma distribution, we merely integrate our truncated PDF for $x\in[0.10,0.24]$ to obtain our CDF.

Let $F$ denote the CDF of a desired finite distribution on some interval $[a,b]$; that is, $F(x)=\mathbb{P}[Y < x]$. In an online fashion, the first coin weight is determined by evaluating 
\[\frac{F(b)-F\left((a+b)/2\right)}{F(b)-F(a)}.\]
This coin is then flipped. If the outcome is $1$, for example, then the next coin weight is determined by evaluating 
\[\frac{F(b)-F\left((a+3b)/4\right)}{F(b)-F\left((a+b)/2\right)}.\] 
Since this process is online, the weight of the next coin flipped depends on the outcome of the previous coin. Based on that outcome, the interval of integration (the evaluation points in the denominator) and the midpoint of the integral (the second evaluation of $F$ in the numerator) change dynamically too.

Compactly, if we let $b$ represent the binary representation of the output random number, we can define the probability that the $k^\text{th}$ bit of $b$ will equal $1$ as
\begin{equation}
\mathbb{P}[b_k=1]=\frac{F(x_2)-F(x_1)}{F(x_2)-F(x_0)},
\end{equation}
where $x_0$ and $x_2$ represent the endpoints of the current interval and $x_1$ represents the midpoint. By dynamically setting $x_0$, $x_1$, and $x_2$ as successive locations according to the outcome of the previous coinflip, we can progressively sample a binary-coded digit from an arbitrary CDF. 

Using this approach, we use a simple recursive algorithm that dynamically sets $x_1$ and $x_2$ depending on the outcome of the previous coinflip $b_k$. If $b_k=1$, then we move the lower bound to the current probability threshold, $x_0\leftarrow x_1$. Likewise, if $b_k=0$, we reset the upper bound ($x_2\leftarrow x_1$). Next, we compute the next as the halfway point between the upper and lower bounds ($x_1 \leftarrow (x_2-x_0)/2$).  This process is illustrated in Fig.~\ref{fig:tree_alg}b.

\section{Device Models: MTJ}\label{Device_Models}

\begin{figure}[htp]
    \centering
    \includegraphics[width=0.96\textwidth]{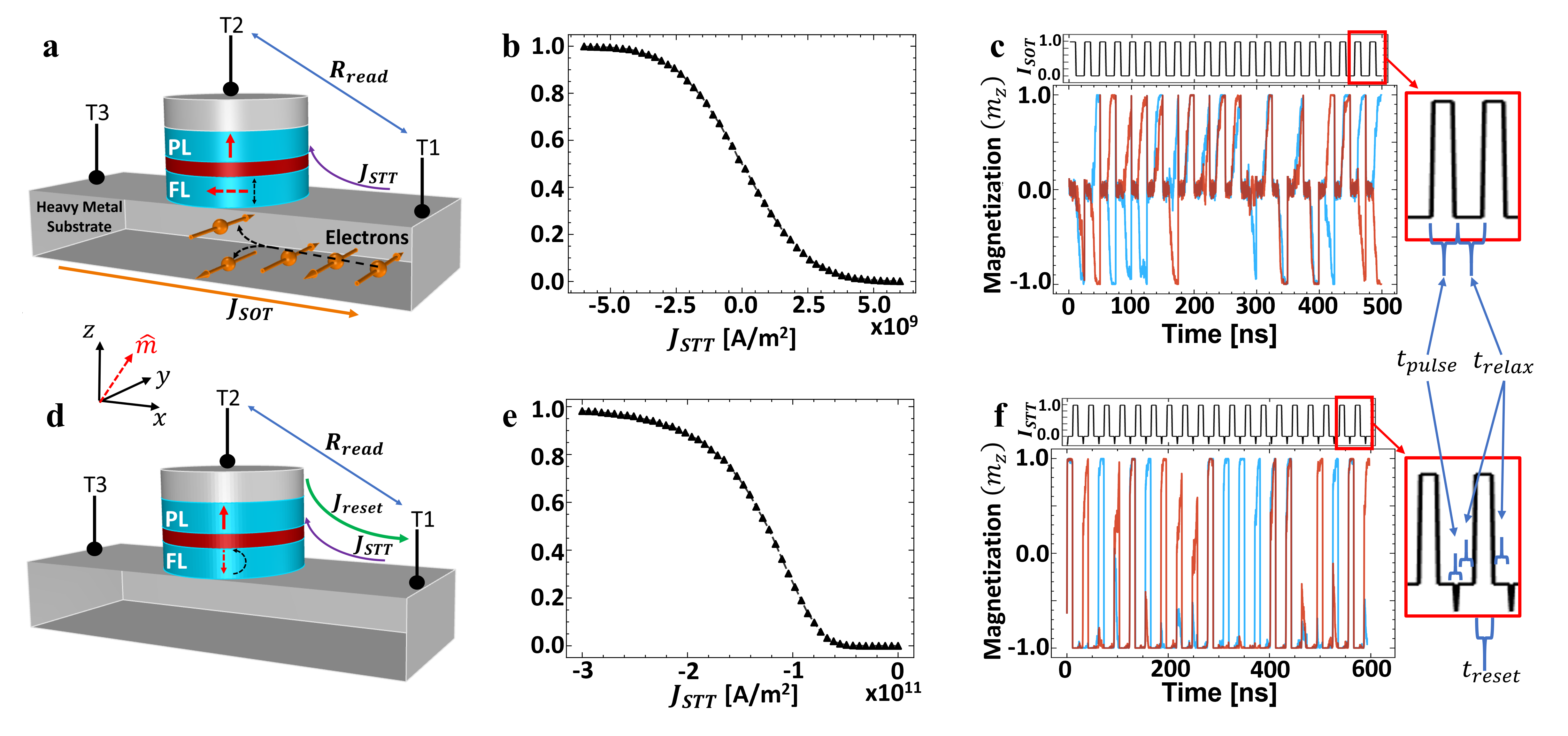} 
    \vskip -0.4 cm
    \caption{Schematic illustrations of the modeled SOT-MTJ (a--c) and STT-MTJ (d--f) with perpendicular magnetic anisotropy. 
    (a) The SOT charge current ($J_{\text{SOT}}$) is applied through terminal T3 to T1 to rotate the free layer (FL) in-plane, and the change of magnetoresistance is read through an MTJ via terminals T2 and T1 after $J_{\text{SOT}}$ is removed. An additional STT charge current $J_{\text{STT}}$ between T1 and T2 provides biasing of the coin. (d) The current ($J_{\text{STT}}$) induces a stochastic switching of the free layer, which is reset with $J_{\text{reset}}$ after each read operation. (b, e) Example device S-curve showing how STT current amplitude ($J_{\text{STT}}$) between T1 and T2 biases the bit probability for both device types. (c, f) Two example random bitstreams generated using the simulation model~\cite{maicke2023magnetic}, with accompanying pulsed SOT or STT currents in units of mA.}
    \label{fig:sot_mtj}
    \vskip -0.3 cm
\end{figure}

The second component of the codesign framework includes the device models, Fig.~\ref{fig:flowchart}a. We utilize physics-based models of these devices to accurately capture the device behaviors based on the device and material properties. This allows us to effectively optimize the devices to target desired distributions while considering additional constraints such as energy efficiency. This component can be updated to explore numerous devices using various modeling techniques, whether it be physics-based models, machine-learning models, or incorporating physical device readings. This flexibility allows the framework to optimize for multiple devices regardless of the modeling approach.

We use a numerical model of MTJs based on a macrospin approximation of the Landau--Lifshitz--Gilbert (LLG) equation~\cite{Wang2015,Leliaert2017}, modeling a standard MTJ stack with PMA comprised CoFeB (free layer)/MgO/CoFeB (fixed layer). This MTJ stack can be operated in different modes for TRNG; an SOT-MTJ device is shown in Fig.~\ref{fig:sot_mtj}a. In the SOT operation, an applied $J_{SOT}$ current in a heavy metal layer beneath the stack rotates the magnetization of the free layer (FL) to be in-plane via the spin Hall effect, then removed for relaxation, as the free layer settles to its PMA lowest energy state to either the 1 or 0 states. This process can be biased using an applied $J_{STT}$ current through the stack. The model is developed in Python and FORTRAN and accounts for device-to-device variability expected for state-of-the-art magnetic random-access memory (MRAM)~\cite{liu2022random,maicke2023magnetic}. An example computed SOT S-curve is shown in Fig.~\ref{fig:sot_mtj}b relating the biasing current $J_{STT}$ to the bit probability, and two example random bitstreams are shown in Fig.~\ref{fig:sot_mtj}c. These bitstreams were produced with extrinsic parameters, applying charge current density of amplitude $J_{\text{SOT}} = \num{-4e11} \,\text{A}/\text{m}^{2}$ without field assist, pulse duration of 10 ns, followed by relaxation of 15 ns. The chosen material parameters for ferromagnetic CoFeB are damping constant $\alpha$  = 0.03, anisotropy constant $K_i$ = $\num{1e-3}\, \text{J}/\text{m}^{2}$, saturation magnetization $M_{s}$ = $\num{1e6}\, \text{A}/\text{m}$, and spin Hall angle $\eta$ = 0.3. Detailed device parameters are shown in Table~\ref{tab:table1}. We leverage this model~\cite{maicke2023magnetic} for optimization experiments at room temperature ($T=300$ K) in the next section to codesign the device as a TRNG for a gamma probability distribution function.

Unlike the SOT-MTJ device, an STT-MTJ device does not require a SOT current to rotate the FL in-plane. Instead, the supplied STT current provides the entire mechanism for stochastic switching. Resetting the FL to be in the $-z$ direction between each sample, an STT current through the stack has a probability to switch the FL depending on the STT magnitude. The STT operation is shown in Fig.~\ref{fig:sot_mtj}d, with corresponding example S-curve and bitstreams in Fig.~\ref{fig:sot_mtj}e,~f. The STT device parameters to generate these bitstreams were the same as those listed for the SOT device, while the pulse, relax, and reset times for the STT are 1 ns, 10 ns, and 10 ns, respectively, and shown in the insets in Fig.~\ref{fig:sot_mtj}c, f.

\begin{table}[ht]
    \tabulinesep=0.6mm
    \begin{tabu}{lccccc}
    \hline
    \multicolumn{6}{|c|}{\textbf{Spin Orbit Torque (SOT) MTJ}} \\ \hline
    \multicolumn{1}{|c|}{\textbf{Parameter}} & \multicolumn{1}{c|}{\textbf{Symbol}} & \multicolumn{1}{c|}{\textbf{Min Range}} & \multicolumn{1}{c|}{\textbf{Max Range}} & \multicolumn{1}{c|}{\textbf{Defaults}} & \multicolumn{1}{c|}{\textbf{Parameter Type}} \\ \hline
    \makecell{Gilbert Damping \\ Constant} & $\alpha$ & 0.01 & 0.1 & 0.03 & Material \\ \hline
    \makecell{Surface Anisotropy \\ Energy} & $K_i$ & \makecell{\num{0.2e-3} \\ $\text{J}/\text{m}^{2}$} & \makecell{\num{1e-3} \\ $\text{J}/\text{m}^{2}$} & \makecell{\num{1e-3} \\ $\text{J}/\text{m}^{2}$} & Device \\ \hline
    \makecell{Saturation \\ Magnetization} & $M_s$ & \makecell{\num{0.3e6} \\ $\text{A}/\text{m}$} & \makecell{\num{2e6} \\ $\text{A}/\text{m}$} & \makecell{\num{1.2e6} \\ $\text{A}/\text{m}$} & Material \\ \hline
    \makecell{Parallel Resistance} & $R_p$ & 500 $\Omega$ & 50000 $\Omega$ & 5000 $\Omega$ & Device \\ \hline
    \makecell{Spin Hall Angle} & $\eta$ & 0.1 & 2 & 0.3 & Material \\ \hline
    \makecell{Current Density} & $J_{\text{SOT}}$ & \makecell{\num{0.01e12} \\ $\text{A}/\text{m}^{2}$} & \makecell{\num{5e12} \\ $\text{A}/\text{m}^{2}$} & \makecell{\num{0.5e12} \\ $\text{A}/\text{m}^{2}$} & Device \\ \hline
    \makecell{Pulse Width} & $t_{\text{pulse}}$ & 0.5 ns & 75 ns & 10 ns & Device \\ \hline
    
    \multicolumn{6}{|c|}{\textbf{Spin Transfer Torque (STT) MTJ}} \\ \hline
    \multicolumn{1}{|c|}{\textbf{Parameter}} & \multicolumn{1}{c|}{\textbf{Symbol}} & \multicolumn{1}{c|}{\textbf{Min Range}} & \multicolumn{1}{c|}{\textbf{Max Range}} & \multicolumn{1}{c|}{\textbf{Defaults}} & \multicolumn{1}{c|}{\textbf{Parameter Type}} \\ \hline
    \makecell{Gilbert Damping \\ Constant} & $\alpha$ & 0.01 & 0.1 & 0.03 & Material \\ \hline
    \makecell{Surface Anisotropy \\ Energy} & $K_i$ & \makecell{\num{0.2e-3} \\ $\text{J}/\text{m}^{2}$} & \makecell{\num{1e-3} \\ $\text{J}/\text{m}^{2}$} & \makecell{\num{1e-3} \\ $\text{J}/\text{m}^{2}$} & Device \\ \hline
    \makecell{Saturation \\ Magnetization} & $M_s$ & \makecell{\num{0.3e6} \\ $\text{A}/\text{m}$} & \makecell{\num{2e6} \\ $\text{A}/\text{m}$} & \makecell{\num{1.2e6} \\ $\text{A}/\text{m}$} & Material \\ \hline
    \makecell{Parallel Resistance} & $R_p$ & 500 $\Omega$ & 50000 $\Omega$ & 5000 $\Omega$ & Device \\ \hline
    \makecell{Pulse Width} & $t_{\text{pulse}}$ & 0.5 ns & 75 ns & 1 ns & Device \\ \hline
    \end{tabu}
    \caption{Device and material parameters for SOT and STT operation, along with their default parameters established through literature and prior works~\cite{cardwell2024, enobio2015cofeb, kateel2023field, garello2018sot, doevenspeck2020sot, cubukcu2014spin}.}
    \label{tab:table1}
\end{table}

\section{AI-Guided Codesign Framework: RL and EA}\label{Optimizations}
The last component of the codesign framework includes the AI-guided codesign strategies, Fig.~\ref{fig:flowchart}b. This work explores and compares both RL and EA to help determine the optimal parameter values for the materials and devices in order to display a desired distribution while accounting for additional constraints, i.e. energy efficiency. The codesign framework is flexible and extendable allowing the investigation of additional constraints to fit the requirements of the given application if needed.

\subsection{RL Approach}
\begin{figure}[htp]
    \centering
    \vskip -0.5cm 
    \includegraphics[width=0.5\textwidth]{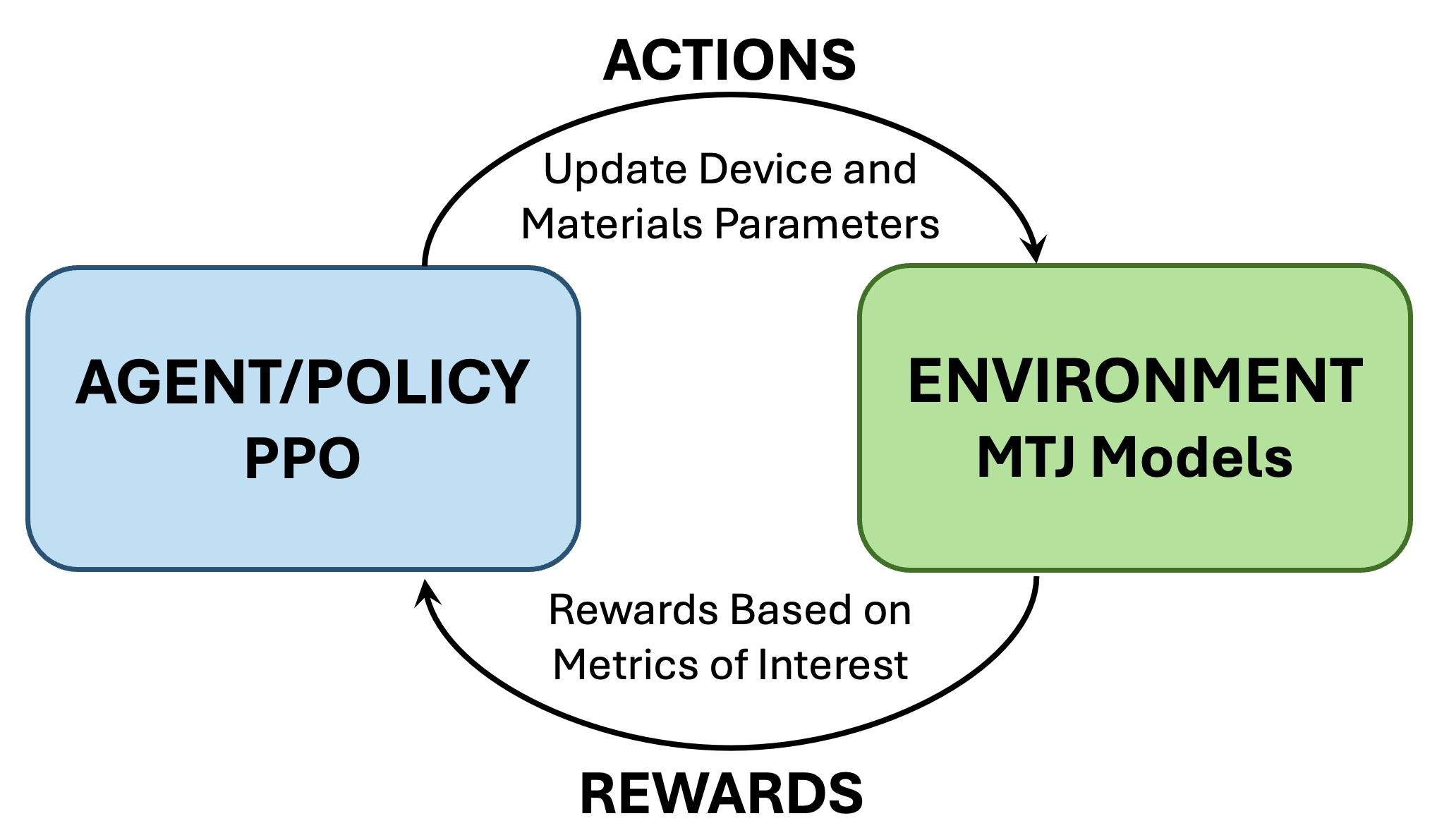} 
    \caption{Overview of RL algorithm for device discovery.}
    \label{fig:RL_diagram}
\end{figure}

RL is a class of machine learning techniques that trains an agent to learn optimal decision making for a given environment (the world the agent inhabits) in order to maximize its rewards~\cite{sutton}. The main control loop begins with the agent taking an action based on the observations available to it for a given state of the environment. The observations provide key information the agent may use to gauge its current situation to determine potential actions to take. The agent is then provided a reward that either penalizes or affirms said action and the environment is updated by transitioning to the next state triggered by the agent's action as shown in Fig.~\ref{fig:RL_diagram} under Supplementary Information.

This state--action--reward dynamic is a crucial component of various RL algorithms. A popular branch of these algorithms is policy-based approaches which is what we are using in this work. This approach focuses on optimizing the policy which is responsible for providing the agent with optimal actions to take. We used a well-established RL algorithm called proximal policy optimization (PPO) by OpenAI for our RL algorithm~\cite{schulman2017proximal}. PPO limits the policy change at each epoch to avoid large policy updates which helps to converge to an optimal solution by mitigating large, destructive policy changes.

For each of the device models solving for the RNG application, a similar RL setup was constructed to train an agent to optimize the materials and parameters of the two device models to best exhibit the desired distribution. To accomplish this, two metrics were leveraged to define optimal configurations: Kullback--Leibler (KL) divergence and energy. KL divergence helps determine how close one distribution is to another, and energy calculations relate to how efficient the device is. Therefore, to obtain the best configurations, the agent is trying to minimize both metrics. For both devices, the agent was trained for 6,000 timesteps and then tested on 150 episodes with each episode consisting of 60 timesteps. For more information, refer to Section~\ref{RL_info} under Supplementary Information.

\subsection{EA Approach}
Evolutionary optimization is an optimization approach inspired by principles in natural evolution. In evolutionary optimization, potential solutions are represented as genomes, and a population of genomes is maintained throughout the optimization. Each genome is converted into a phenome and evaluated through a fitness function, which corresponds to the objective function that is either minimized or maximized. Selection procedures are used to select parents from the population of individuals based on their fitness values. Then, reproduction operations such as random mutation are performed with the parents to produce children genomes, which then replace the parents. To allow for multi-objective optimization, we use the non-dominated sorting genetic algorithm II (NSGA-II)~\cite{deb2002fast}. In NSGA-II, individuals are selected to fall along the Pareto front, with a crowding distance metric that is used to promote diversity among the selected individuals.

Here, the individual genomes in our population are real-valued arrays, where each gene in the genome represents the parameters of the device type. We use Gaussian mutations to mutate the real values in the array. In evaluating the device performance, we use two objectives: KL divergence and energy usage. The goal is to minimize both of those values---similar to the RL approach. If the configuration is invalid, then the KL divergence and energy usage values are set to a large value (1,000,000) in order to strongly discourage invalid configurations. For both devices, 25 runs were performed with a population size of 50 for 50 generations of evolution.

\section{AI-Guided Optimization Results}\label{Results}

\subsection{RL Results}

\captionsetup[subfigure]{labelfont=bf, labelformat=simple, singlelinecheck=false}
\begin{figure}[htp]
    \centering
    \begin{subfigure}{0.49\textwidth}
        \caption{}
        \includegraphics[width=\textwidth]{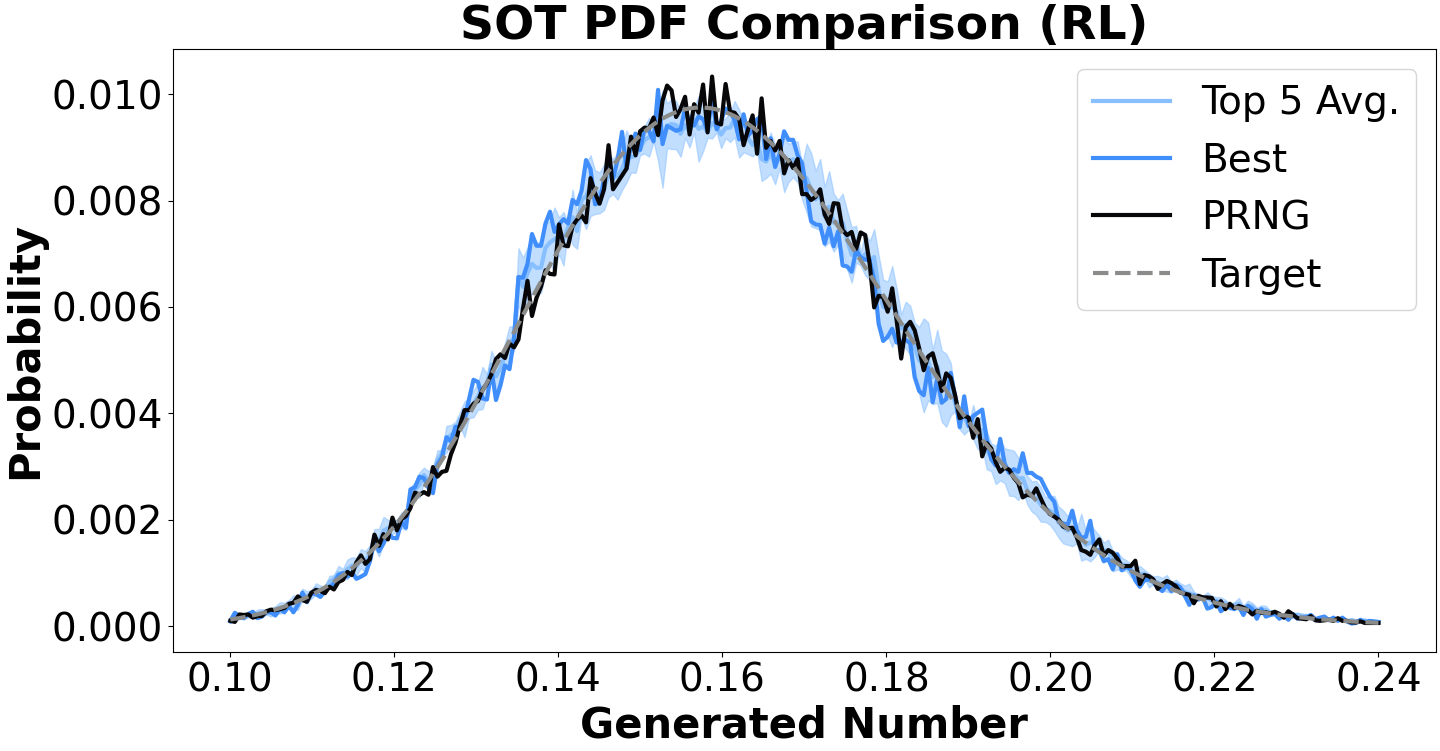}
        \label{fig:rl_sot_pdf}
    \end{subfigure}
    \hfill
    \vspace{\floatsep}
    \begin{subfigure}{0.49\textwidth}
        \caption{}
        \includegraphics[width=\textwidth]{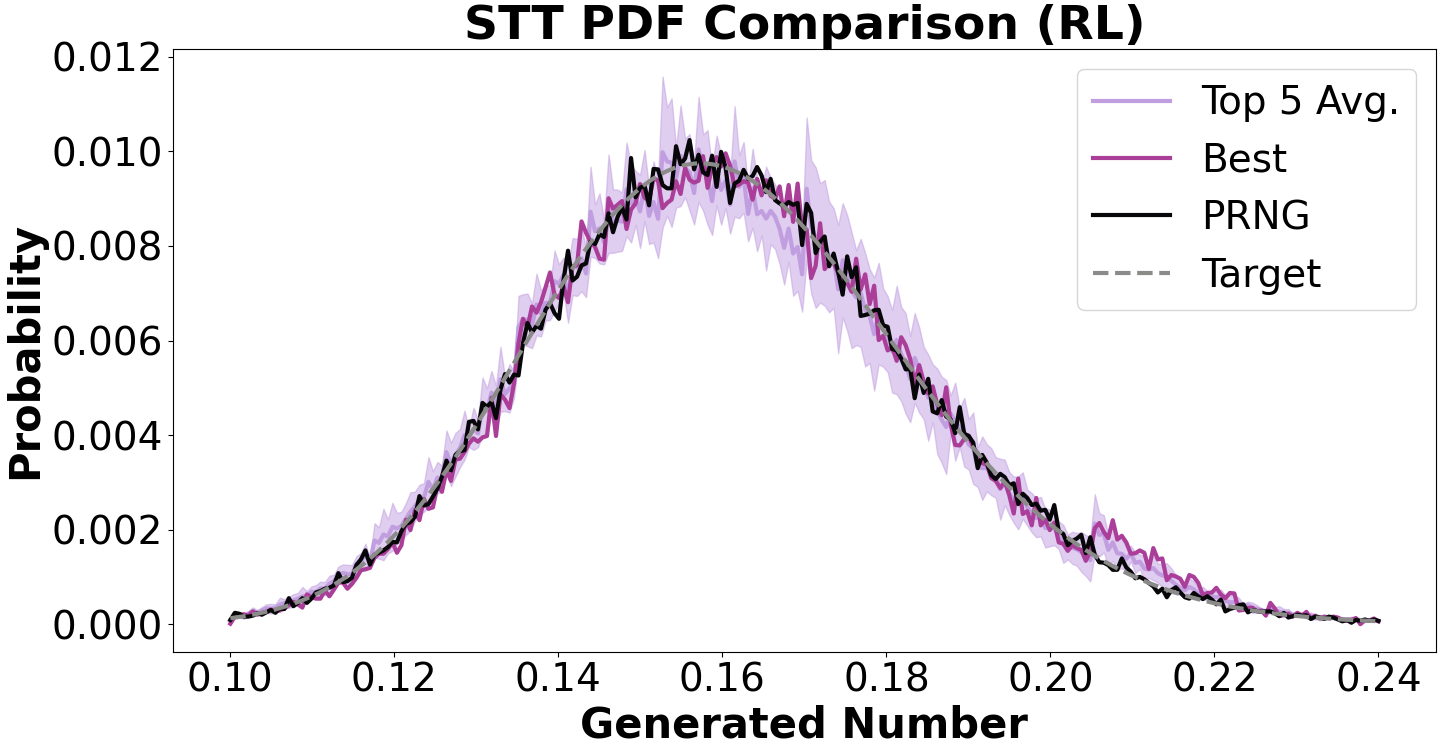}
        \label{fig:rl_stt_pdf}
    \end{subfigure}
    \vskip -1 cm
    
    \begin{subfigure}{0.49\textwidth}
        \caption{}
        \includegraphics[width=\textwidth]{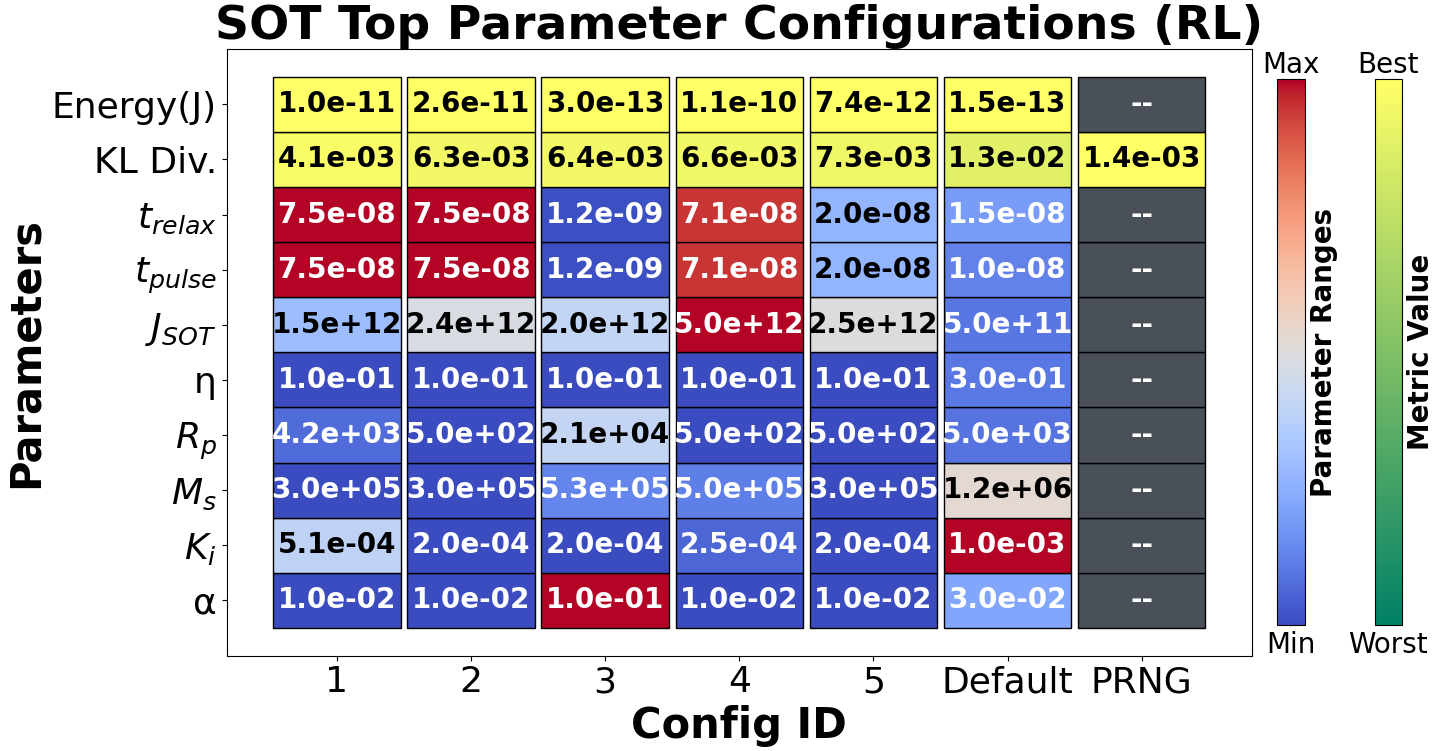}
        \label{fig:rl_sot_params}
    \end{subfigure}
    \hfill
    \vspace{\floatsep}
    \begin{subfigure}{0.49\textwidth}
        \caption{}
        \includegraphics[width=\textwidth]{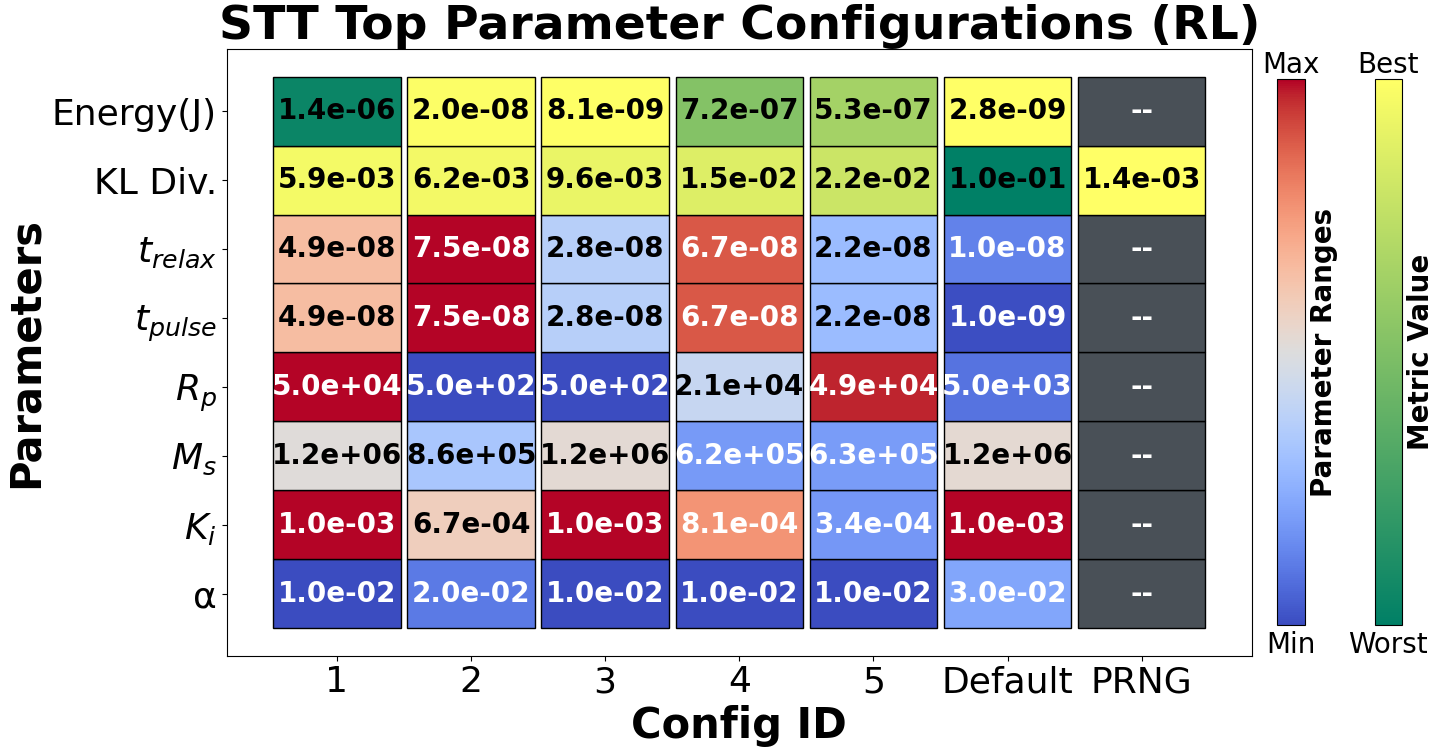}
        \label{fig:rl_stt_params}
    \end{subfigure}
    \vskip -1 cm
    
    \begin{subfigure}{0.49\textwidth}
        \caption{}
        \includegraphics[width=\textwidth]{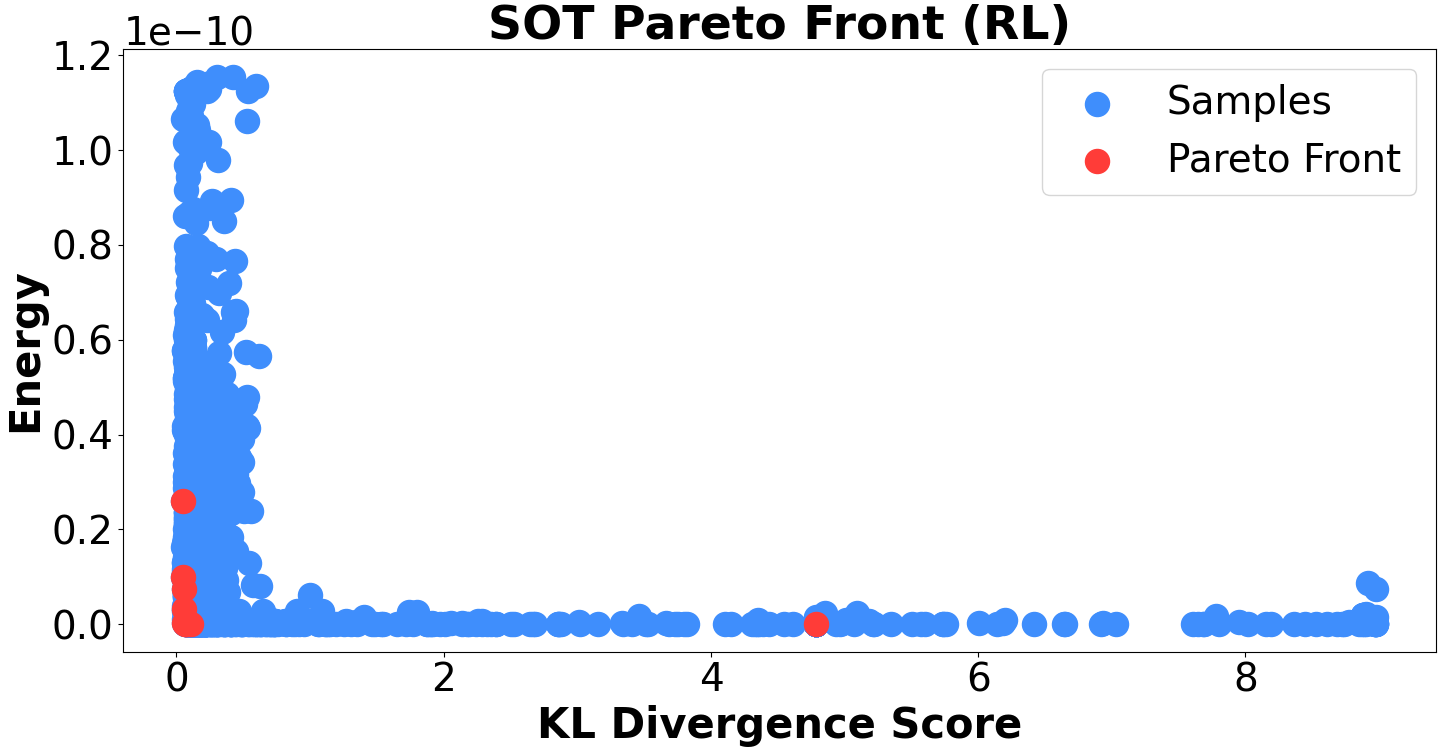}
        \label{fig:rl_sot_pareto}
    \end{subfigure}
    \hfill
    \vspace{\floatsep}
    \begin{subfigure}{0.49\textwidth}
        \caption{}
        \includegraphics[width=\textwidth]{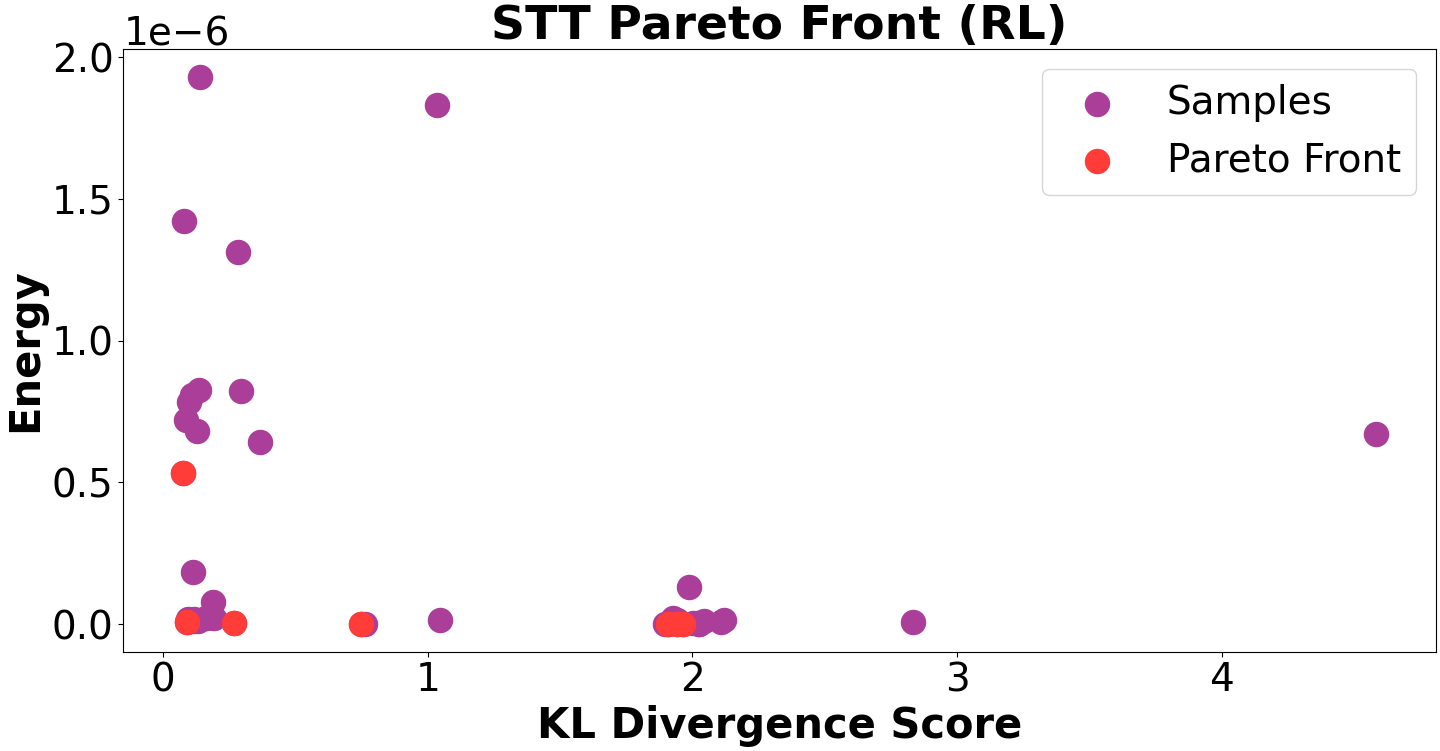}
        \label{fig:rl_stt_pareto}
    \end{subfigure}
    \vskip -1 cm

    \begin{subfigure}{0.49\textwidth}
        \caption{}
        \includegraphics[width=\textwidth]{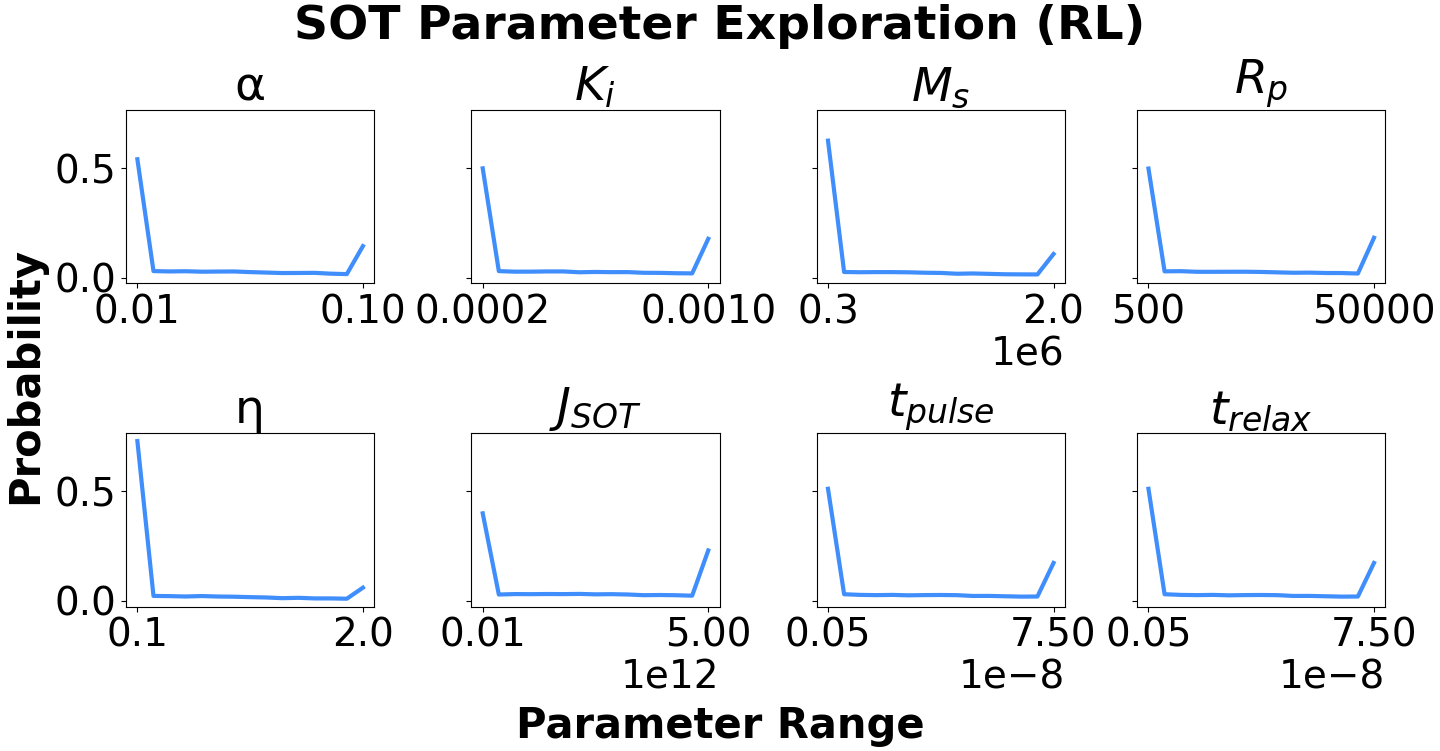}
        \label{fig:rl_sot_exploration} 
    \end{subfigure}
    \hfill
    \vspace{\floatsep}
    \begin{subfigure}{0.49\textwidth}
        \caption{}
        \includegraphics[width=\textwidth]{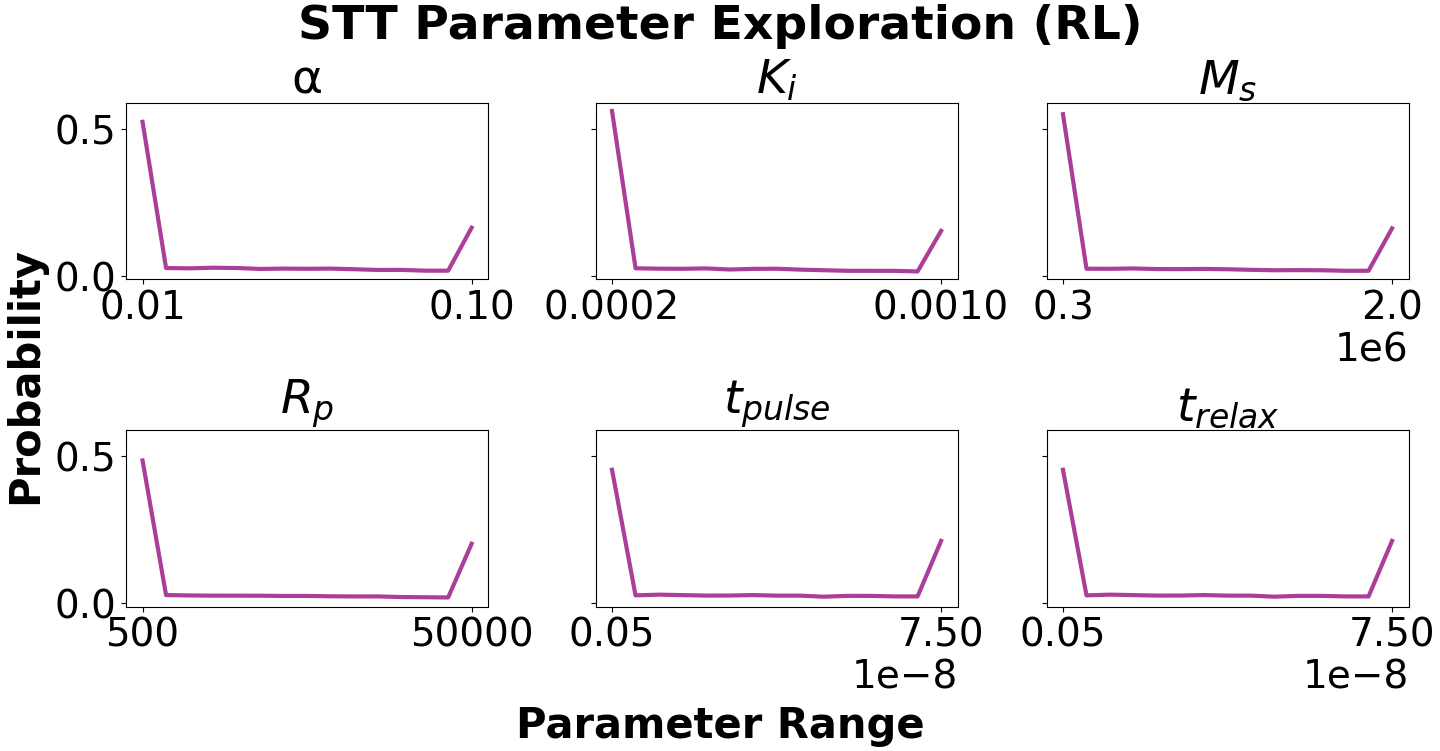}
        \label{fig:rl_stt_exploration}
    \end{subfigure}
    \vskip -1 cm
    
\caption{RL device optimization results. (a, b) PDF comparison of top 5 device configurations, best configuration, PRNG, and target distribution for both SOT and STT devices. (c, d) Parameter configurations of top 5 devices with energy and KL divergence metrics compared against the default configurations and a PRNG for both SOT and STT devices. (e, f) Pareto fronts comparing energy and KL divergence metrics of various SOT and STT device configurations. (g, h) Probability distributions of the parameter ranges that were explored for both SOT and STT devices.}
\label{fig:rl_results}
\end{figure}

Fig.~\ref{fig:rl_results} shows an overview of the results for our RL approach to device design. Fig.~\ref{fig:rl_sot_pdf},~b show the PDF comparisons for the SOT and STT devices, respectively. Each figure shows the distributions produced by the average of the top 5 devices generated through RL, the distribution produced by the best device configuration, and the result of the PRNG all compared against the target distribution. Note that the PRNG was sampled the same amount as the MTJ devices (100,000) to allow for comparisons. As can be seen in these figures, the optimized device results match well to the target distribution and are mostly consistent with what is seen for the PRNG. We can see in Fig.~\ref{fig:rl_stt_pdf} that there is significantly more variation for the STT device than the SOT device, particularly with drift from the distribution of the generated numbers between 0.20 and 0.22. This is expected behavior, as the STT device stochasticity should be harder to tune due to having one control knob while the SOT device has two control knobs. While the SOT device needs its free layer to be brought in-plane and then allowed to relax, the current to flip the STT device needs to be fairly precise to achieve proper biasing. However, the tradeoff is added complexity and terminals to the SOT-MTJ compared to the STT version that can use a 1 transistor, 1 resistor STT-MRAM structure.

Fig.~\ref{fig:rl_sot_params},~d show the five best configurations discovered for the SOT and STT devices, respectively, along with their respective KL divergence and energy metrics. The top 5 devices are considered to allow comparison of configuration variability through the optimization strategies among the top performers. In order to provide a baseline, these optimized configurations are compared against a PRNG and ``default'' values for each device parameter, established through literature and prior works~\cite{enobio2015cofeb, kateel2023field, garello2018sot, doevenspeck2020sot, cubukcu2014spin}. It is worth noting that the individual color ranges for the energy and KL divergence metrics were determined by looking at all device configurations (SOT and STT), for both automated codesign strategies (RL and EA), including the default configurations and the PRNG. The optimized configurations for both devices were able to outperform the default configurations in terms of KL divergence. However, the energy metric wasn't improved upon from the default configuration, and this can be attributed to less weight being applied to the energy metric in the fitness score---increasing the weight can incentivize better energy efficiency. It is clear that there is not one set of parameters that leads to the best performance for a given device set, though there are trends for certain parameters. In particular, the Gilbert damping constant ($\alpha$) is consistent across both the STT and SOT devices. 
For the SOT device, the $t_{\text{relax}}$ and $t_{\text{pulse}}$ parameters tend to stay toward the high end of the allowed range, but the same is not true for the STT device, where there is more variation in the parameter values used in the top 5 devices. This indicates that it is worthwhile to customize the parameters for each individual device type. For the remaining parameters, values across the ranges are used for each of the top 5 device configurations for the STT device, but values toward the lower bound are preferred for the SOT device which may suggest the agent falling in a local optima. Looking at the energy and KL divergence metrics, we can see that the SOT device produced more energy-efficient configurations that more closely matched the target distribution compared to the STT device.

Fig.~\ref{fig:rl_sot_pareto},~f depict the Pareto fronts comparing energy consumption and KL divergence of the valid configurations for both SOT and STT devices, respectively. As seen by the figures, there are significantly more samples for the SOT device compared to the STT device. This suggests that there is a smaller subset of valid configurations for the STT device as opposed to the SOT device, as expected. 
This disparity allowed for a larger exploration of the energy and KL divergence space for the SOT device, while the STT device has clusters of samples, suggesting that there may be fewer configurations that deviate too far from the desired distribution as seen by fewer occurrences of larger KL divergence scores. Nevertheless, both devices share a similar trend: closely matching the target distribution (lower KL divergence score) typically results in increased energy consumption. This tradeoff is a crucial element to be aware of when designing such devices since it offers a gradient of possibilities to best match the application requirements.

Fig.~\ref{fig:rl_sot_exploration},~h showcase the probability distributions of the parameter ranges that were explored through RL for both SOT and STT devices, respectively. From the graphs, we can clearly see that the majority of the parameter exploration took place at the bounds of the parameter ranges for both devices. This matches with the top parameters discovered (Fig.~\ref{fig:rl_sot_params},~d), where the parameter values tend to be toward the bounds of their respective ranges. This suggests that the agent may have fallen into a local optimum; however, the Pareto fronts suggest that the agent was still able to discover a wide breadth of the performance metrics' (energy and KL divergence) space.

\subsection{EA Results}

\begin{figure}[htp]
    \centering
    \begin{subfigure}{0.49\textwidth}
        \caption{}
        \includegraphics[width=\textwidth]{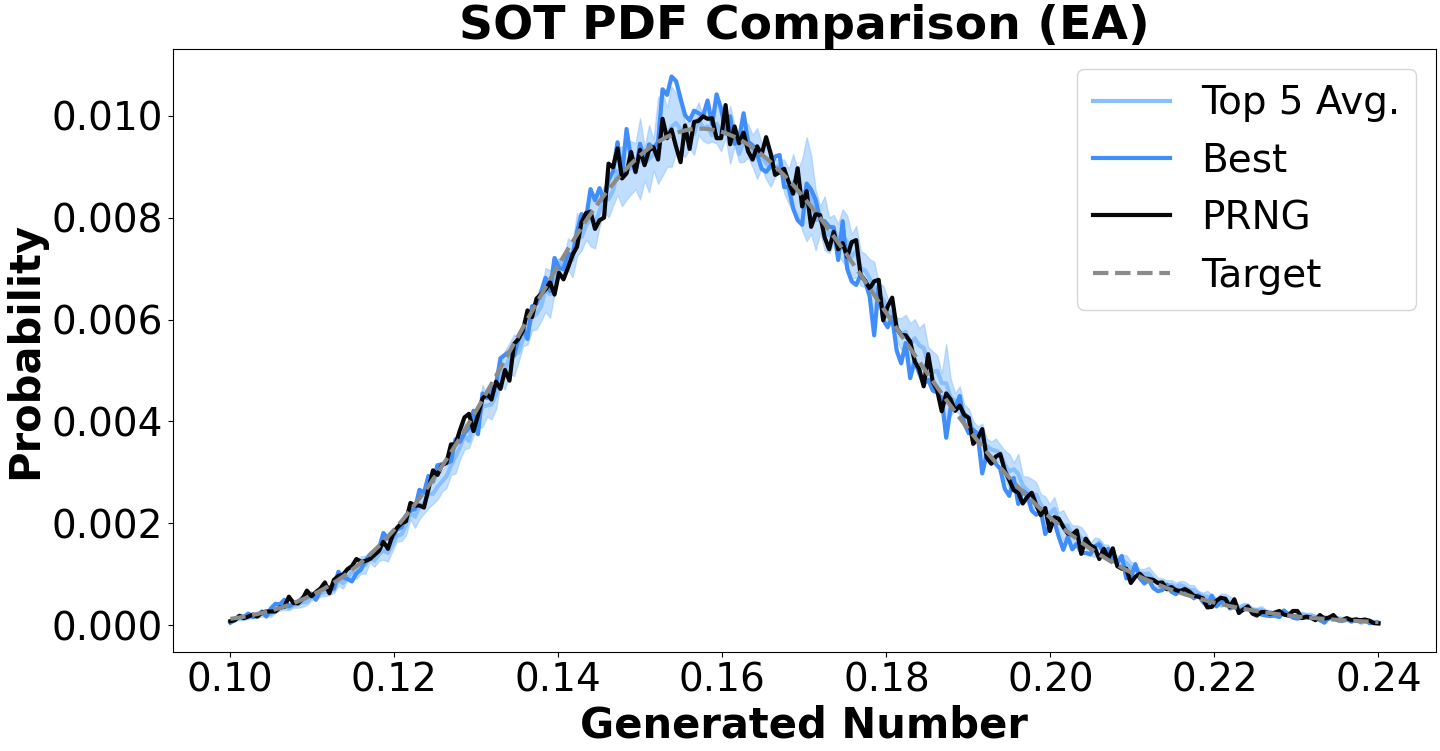}
        \label{fig:leap_sot_pdf}
    \end{subfigure}
    \hfill 
    \begin{subfigure}{0.49\textwidth}
        \caption{}
        \includegraphics[width=\textwidth]{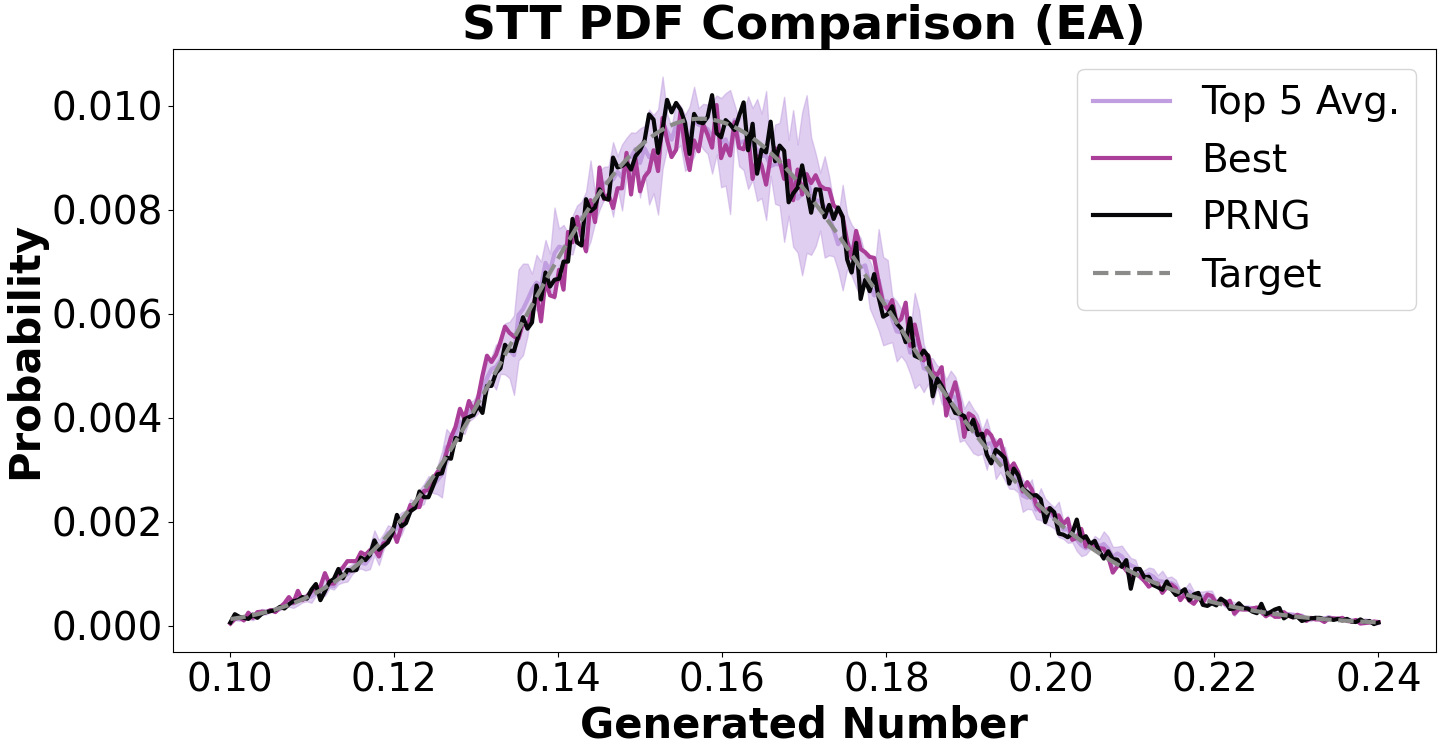}
        \label{fig:leap_stt_pdf}
    \end{subfigure}
    \vskip -1 cm

    \begin{subfigure}{0.49\textwidth}
        \caption{}
        \includegraphics[width=\textwidth]{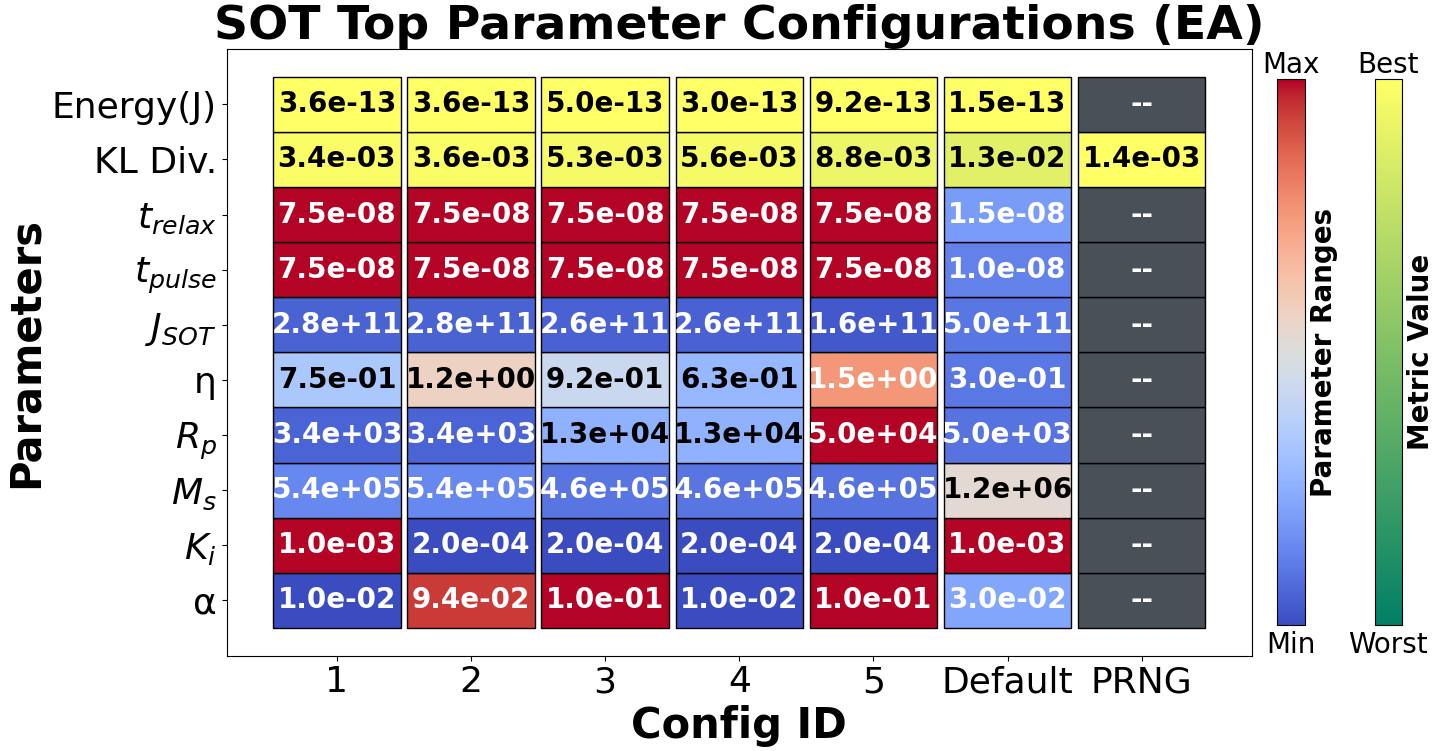}
        \label{fig:leap_sot_params}
    \end{subfigure}
    \hfill
    \begin{subfigure}{0.49\textwidth}
        \caption{}
        \includegraphics[width=\textwidth]{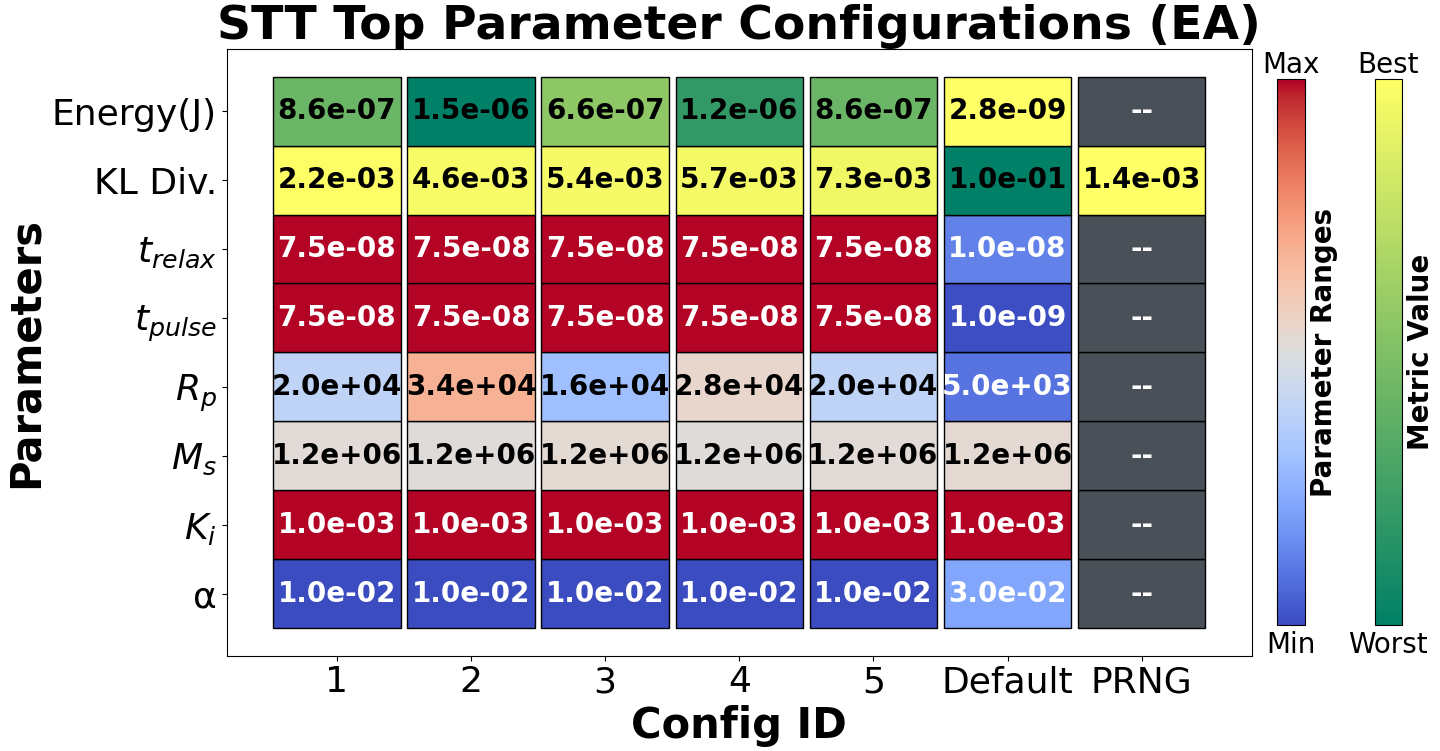}
        \label{fig:leap_stt_params}
    \end{subfigure}
    \vskip -1 cm

    \begin{subfigure}{0.49\textwidth}
        \caption{}
        \includegraphics[width=\textwidth]{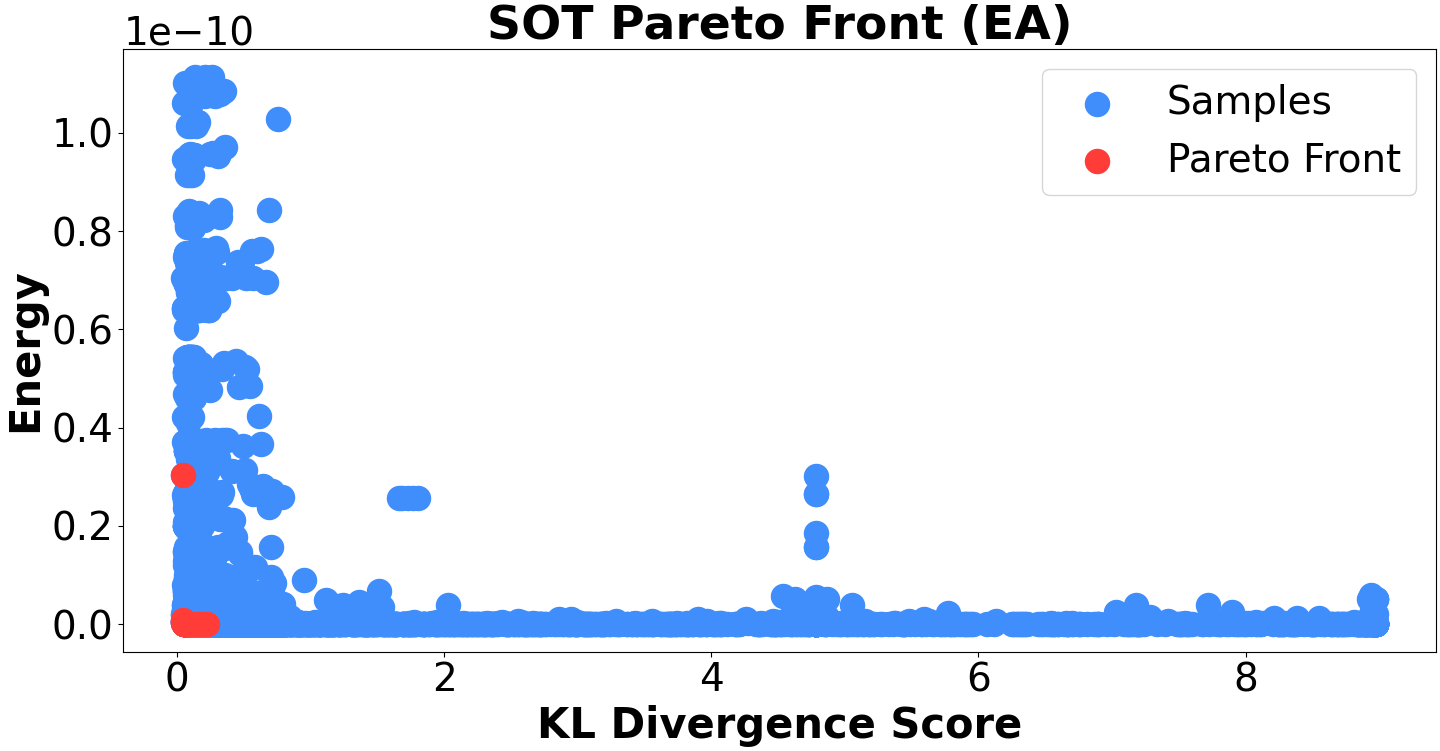}
        \label{fig:leap_sot_pareto}
    \end{subfigure}
    \hfill
    \begin{subfigure}{0.49\textwidth}
        \caption{}
        \includegraphics[width=\textwidth]{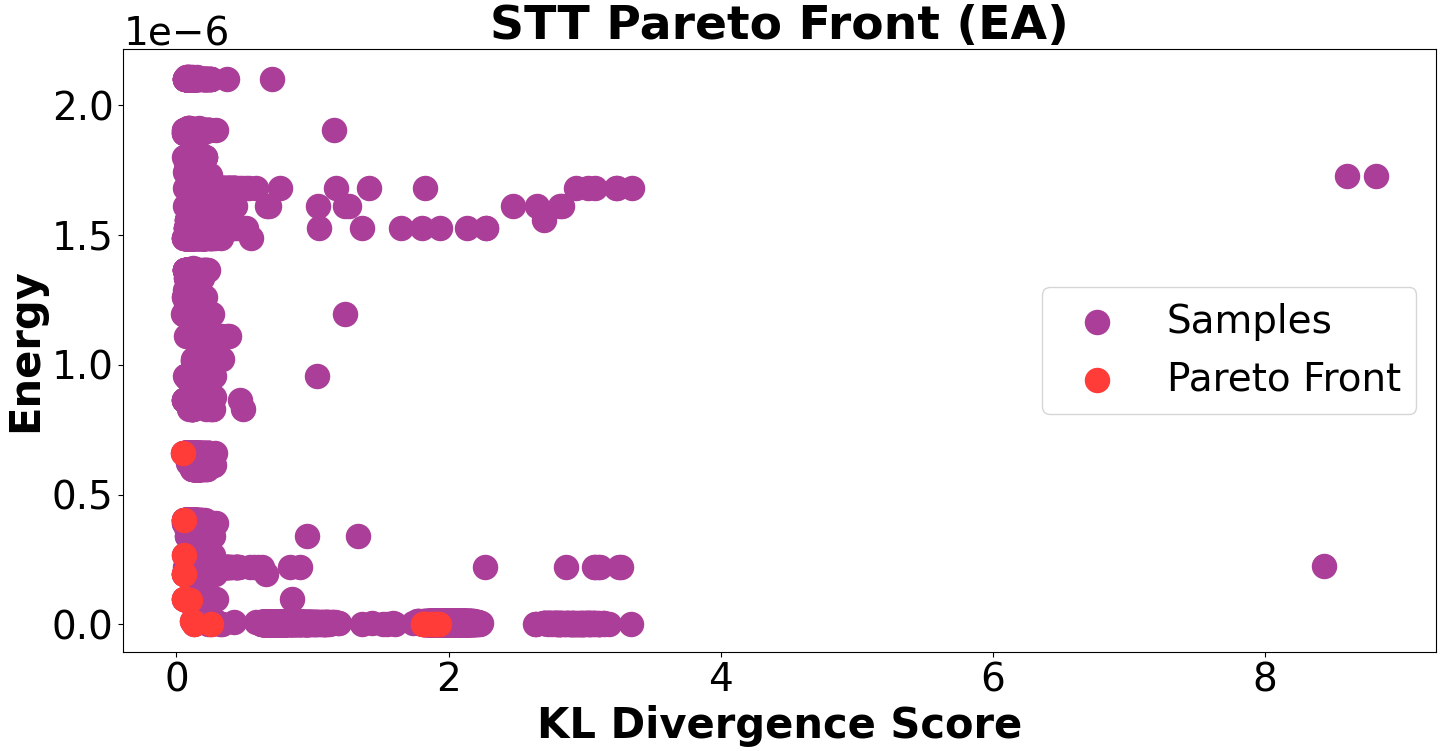}
        \label{fig:leap_stt_pareto}
    \end{subfigure}
    \vskip -1 cm

    \begin{subfigure}{0.49\textwidth}
        \caption{}
        \includegraphics[width=\textwidth]{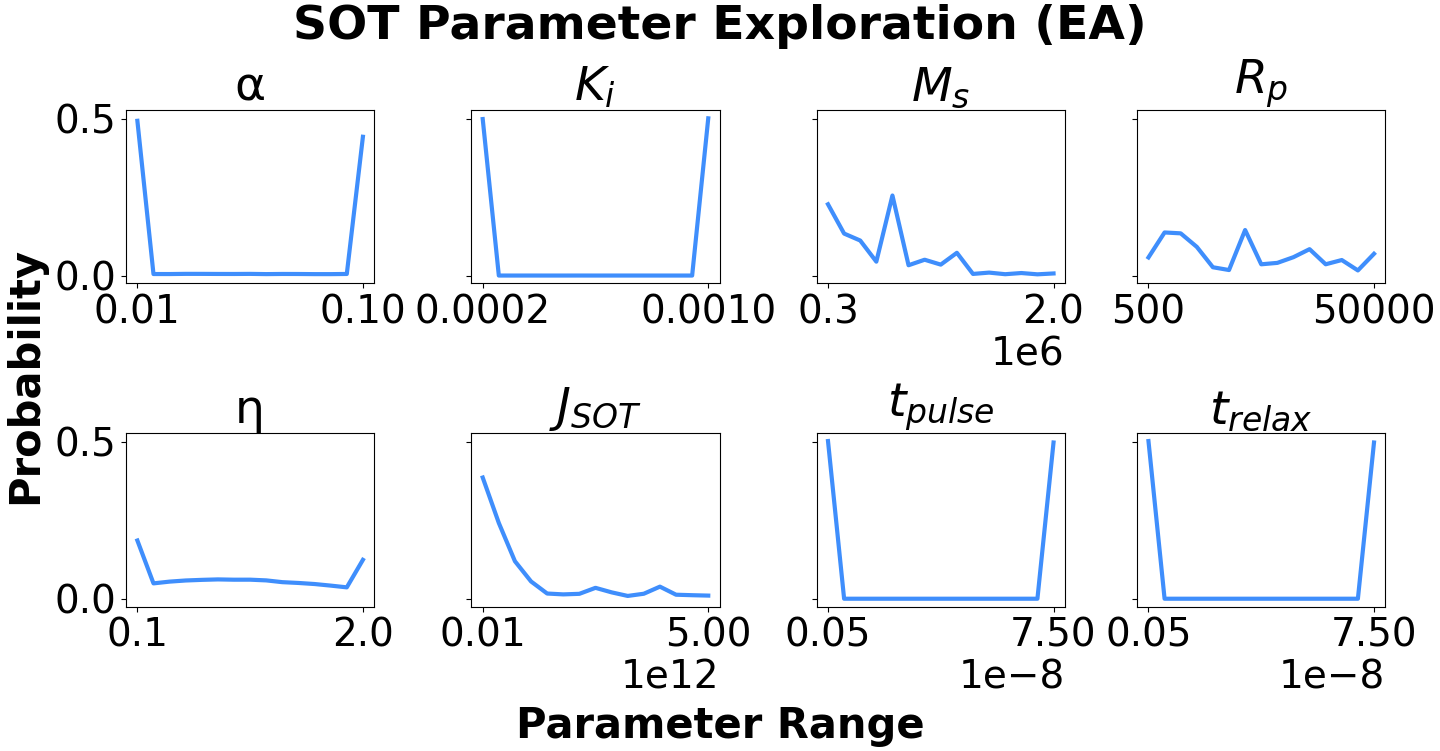}
        \label{fig:leap_sot_exploration}
    \end{subfigure}
    \hfill
    \begin{subfigure}{0.49\textwidth}
        \caption{}
        \includegraphics[width=\textwidth]{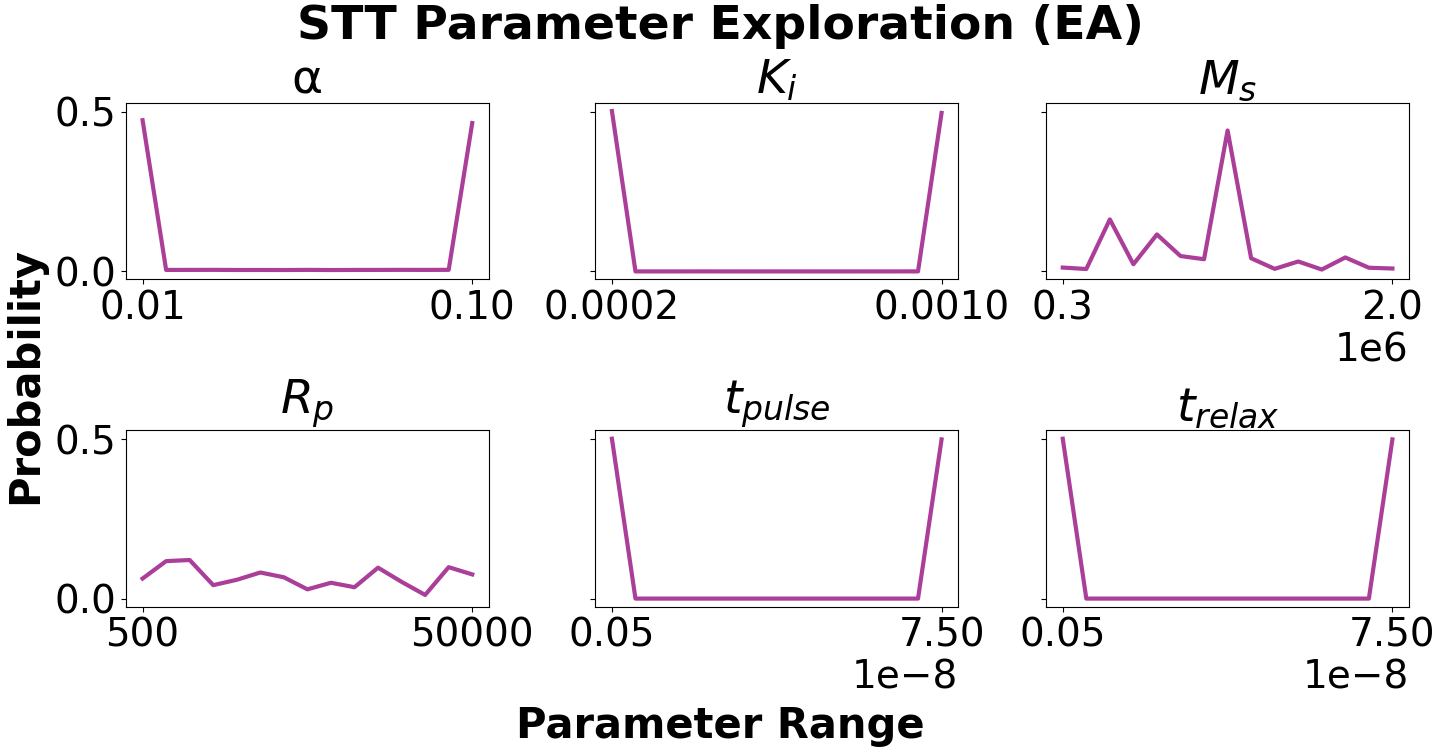}
        \label{fig:leap_stt_exploration}
    \end{subfigure}
    \vskip -0.5 cm

\caption{EA device optimization results. (a, b) PDF comparison of top 5 device configurations, best configuration, PRNG, and target distribution for both SOT and STT devices. (c, d) Parameter configurations of top 5 devices with energy and KL divergence metrics compared against the default configurations and a PRNG for both SOT and STT devices. (e, f) Pareto fronts comparing energy and KL divergence metrics of various SOT and STT device configurations. (g, h) Probability distributions of the parameter ranges that were explored for both SOT and STT devices.}
\label{fig:leap_results}
\end{figure}

Fig.~\ref{fig:leap_results} provides an overview of the EA results for device design. Fig.~\ref{fig:leap_sot_pdf},~b display the PDF comparisons of the SOT and STT devices, respectively. Once again, we see that the device parameters discovered by the EA closely mimic the desired distribution and the distribution generated by the PRNG. Similarly to the RL results, we also see that the STT device has more variation from the distribution than the SOT device, but this time for generated numbers between 0.16 and 0.18, which indicates that different parameter sets may result in more variation at different points in the distribution.

Fig.~\ref{fig:leap_sot_params},~d show the top 5 device configurations discovered across EA runs along with their KL divergence and energy metrics for the SOT and STT devices, respectively. These are compared against the default configurations and PRNG. Similar to the RL results, these configurations were also able to outperform the default configurations in terms of KL divergence. However, the energy metric is relatively close to the default configuration for the SOT device, except for the STT device, where the optimized configurations had better KL divergence scores at the expense of increased energy consumption compared to the default configuration. Additionally, the device configurations are much more similar to each other, with consistent values discovered for $t_{\text{relax}}$, $t_{\text{pulse}}$, and $\alpha$ across a majority of the best configurations for both device sets. This is most prominently seen in the STT device, where most of the variation occurs in the $R_{\text{p}}$ parameter. The parameter values also tend to be closer to the bounds of the respective ranges except for the $R_{\text{p}}$ parameters for STT and the $\eta$ parameter for SOT.

Fig.~\ref{fig:leap_sot_pareto},~f depict the Pareto fronts for the multi-objective EA, comparing energy consumption and KL divergence of the valid configurations for both SOT and STT devices, respectively. We can see that there are a variety of device types that are explored across the performance space for both SOT and STT. These figures once again show that there are many configurations that use less energy but have high KL divergence. Unlike the RL results, Fig.~\ref{fig:leap_stt_pareto} shows more exploration in the STT space. This plot shows that the EA likely explored two distinct local optima in the search space at two different energy levels (one higher and one lower) with varying levels of KL divergence. However, well-performing (low KL divergence and low energy) device configurations were still discovered. Similar to RL, there are fewer valid configurations for the STT device, which further supports the idea that there is a smaller subset of valid configurations for this device type.

Fig.~\ref{fig:leap_sot_exploration},~h show the probability distributions of the parameter ranges that were explored through the EA approach for both SOT and STT devices, respectively. We can see in these plots that only the extremes in the parameter range are explored for some parameter sets like $\alpha$, $t_{\text{pulse}}$, $t_{\text{relax}}$, and $K_{\text{i}}$. This is consistent with what we saw for those parameter values in the RL approach as well. However, while the RL approach kept to the extremes for all of the device parameter values, the EA explores across the ranges for some of the device parameters, in particular $M_{\text{s}}$ and $R_{\text{p}}$. We can confirm in Fig.~\ref{fig:leap_sot_params},~d that the best-performing configurations for $R_{\text{p}}$ and $M_{\text{s}}$ tended to fall more toward the middle of the available parameter range, rather than at one of the two extremes.

\section{Discussion}\label{Discussion}

The key contribution of this work is the complete device codesign framework for true random number generation based on abstract device models for EA and RL approaches, which is illustrated in Fig.~\ref{fig:flowchart}. The creation of such a framework requires device expertise, AI expertise, and application expertise. However, now that the framework and components have been established, it is worth noting that though we have showcased results for EA- and RL-based automated codesign approaches, SOT and STT-MTJ devices, and gamma distributions, we can easily extend this work in the future by augmenting or swapping out components within the framework. First, we can investigate other design or optimization approaches for the device parameters, such as particle swarm optimization or generative AI. Next, we can swap out or augment the MTJ devices with other probabilistic devices. Finally, we can use the framework to pick materials and device parameters for other non-uniform distributions or target codesign for other application domains. 

Our second contribution is the demonstration of codesigned materials and device parameters for both SOT and STT-MTJ devices for efficient random number generation and thus, new candidate devices. It is important to note that though the default parameter values for both the SOT and STT devices are very similar, both RL and EA approaches discovered parameter sets that were clearly different for each of the two devices. This clearly indicates that leveraging an automated optimization approach for discovering device parameters is worthwhile for these sets of devices. Specifically, by leveraging an optimization approach for each individual device type and application combination, device parameter sets can be discovered that improve performance over the defaults for a particular application, while also tuning the parameters to incentivize certain metrics such as energy efficiency. In this work, the parameter ranges were kept in ranges realistically achievable in the CoFeB--MgO system through materials engineering; future work could expand the ranges further to explore not yet discovered systems.

The last main contribution of this work is the comparison of RL and EA approaches for device design. In comparing these two approaches, there are several points worth noting. First, the EA approach we used, NSGA-II, is particularly focused on multi-objective optimization and exploration of points along the Pareto front, whereas we used a reward function that weighted the objectives into a single score for the RL approach. It is worth noting that although multi-objective approaches exist for RL~\cite{hayes2022practical}, they are less popularly used than EA algorithms such as NSGA-II, which is why we focused on a weighted reward function for RL here. However, determining the appropriate weights for a multi-objective weighted reward function is non-trivial and often requires trial-and-error to determine the best weights for each objective.

We observe that both EA and RL explored configurations across the Pareto fronts; however, the EA approach explored the parameter search space to a larger degree, while the RL approach tended to stay toward the extremes. It is worth noting that this can be at least partially attributed to the smaller number of training timesteps (6,000) for the RL approach compared to the EA approach ($50 \text{ population size} * 50 \text{ generations} * 25 \text{ runs} = 62,500 \text{ timesteps}$). This disparity between the number of training timesteps for the two approaches was due to time and computing constraints; therefore, we would like to revisit this point in the future to compare the two approaches with similar training times. However, even with a smaller number of training timesteps, the RL approach was able to produce more unique configurations in its top-performing candidates compared to the EA approach as demonstrated by Fig.~\ref{fig:rl_sot_params},~d and Fig.~\ref{fig:leap_sot_params},~d. Another observation is that both optimization strategies resulted in the STT device configurations displaying larger variations in their PDFs compared to the SOT device. This, once again, suggests that the STT device has a smaller subset of valid configurations and may be more sensitive to the parameter configurations. Overall, both RL and EA approaches have clear advantages and disadvantages.

\begin{figure}[htp]
    \centering
    \includegraphics[width=0.96\textwidth]{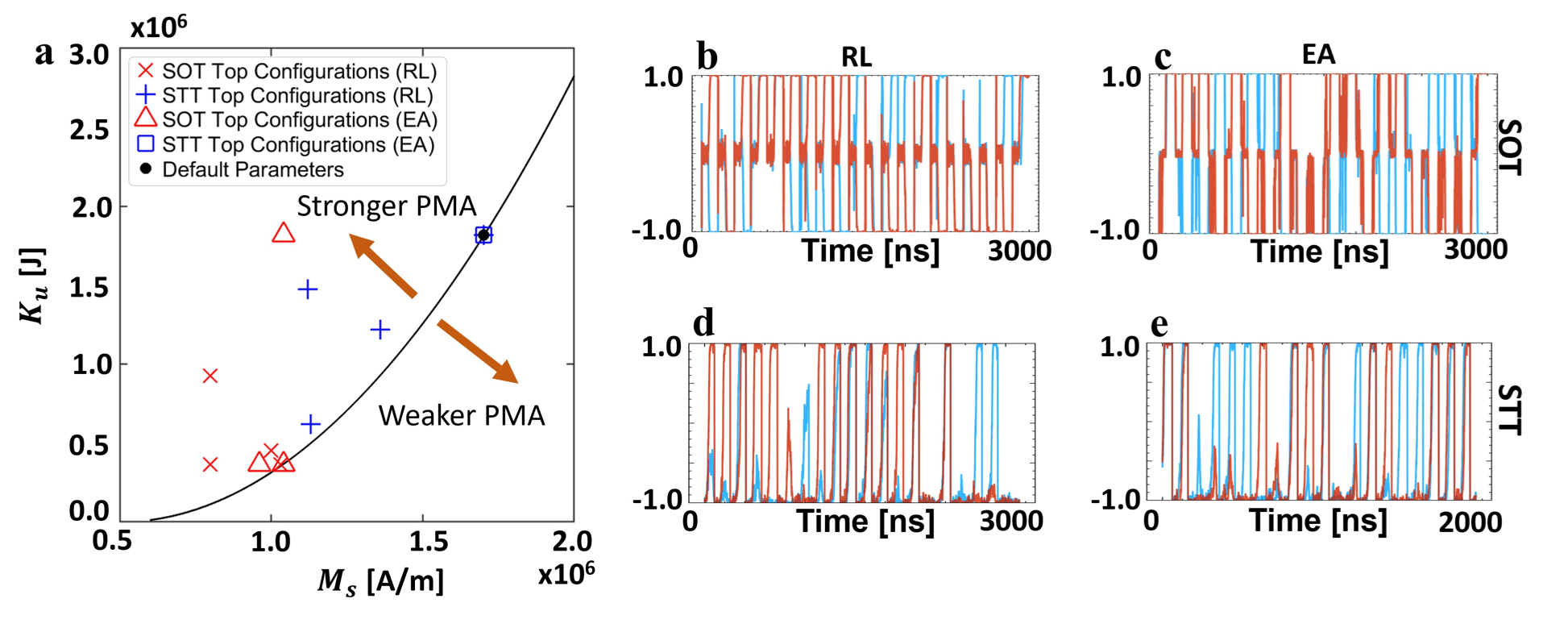} 
    \vskip -0.4 cm
    \caption{ (a) The $M_s$ and $K_u$ pairs for the top five configurations of each of the SOT and STT devices found via RL and EA approaches. The SOT and STT devices generally favor configurations that have stronger PMA, though the SOT devices favor weaker $M_s$ than the STT devices. (b, c) Sample bitstreams generated using the top SOT configuration found by RL and EA approaches. (d, e) Sample bitstreams generated using the top STT configuration found by RL and EA approaches.}
    \label{fig:ms_ki}
    \vskip -0.3 cm
\end{figure}

We summarize the $K_u$ ($K_u = K_i t_f$) and $M_s$ pairs found by each approach for each device in Fig.~\ref{fig:ms_ki}. To orient the configurations within the plot, we derive a simple relationship between $K_u$ and $M_s$ to achieve PMA. Considering only the anisotropy energy density, the total effective anisotropy can be written as
\[K_{eff} = K_u -\frac{\mu_0 M_s^2}{2}.\]
If $K_{eff} > 0$, the device will exhibit stronger PMA behavior, while if $K_{eff} < 0$ the PMA will weaken (and eventually become IMA). When  $K_{eff} = 0$, we can define a `border' by $K_u=\frac{\mu_0 M_s^2}{2}$; this line is shown in black in Fig.~\ref{fig:ms_ki}a. Looking at the regions inhabited by each device, the STT devices remain relatively close to the border, while the SOT devices favor stronger PMA. This is likely because a strong PMA is critical for the SOT device to return to a $\pm z$ magnetization state during relaxation. Example bitstreams generated by SOT devices found via the RL and EA approaches are shown in Fig.~\ref{fig:ms_ki}b,~c, showing the expected SOT behavior. For most SOT devices in the configurations list the bitstreams exhibit this desired functionality, however, it is noted that occasionally the optimization strategies cause the SOT devices to operate in a stochastic regime instead, including the top discovered configurations; see Supplementary section 9.3 for more information. Further, the top STT device bitstreams in Fig.~\ref{fig:ms_ki}d,~e look as expected, stochastically switching to $+z$ with an application of the STT current. Finally, it is noted that the tendency of both approaches to minimize $\alpha$ is likely related to the critical current required to drive MTJ switching; in~\cite{Tatara2004}, it was shown that the critical switching current depends proportionally on $\alpha$, and thus a smaller $\alpha$ allows for switching to occur at lower current densities.

\section{Conclusions}\label{Conclusion}

In this work, we proposed an AI-guided codesign framework to optimize SOT-MTJ and STT-MTJ devices for a TRNG application by utilizing RL and EA. We successfully motivated such a framework by producing valid device configurations that closely match the target distribution while optimizing for energy efficiency. We also demonstrate the feasibility of swapping components of the workflow to fit application requirements by switching out different device types (SOT vs. STT) and design/optimization strategies (RL vs. EA). We found that by optimizing the parameters of the devices beyond their default values, we can tune the MTJ devices to more closely mimic the desired distribution while simultaneously minimizing energy usage, demonstrating that using optimization-guided codesign approaches can discover or optimize devices for particular applications.

As noted previously, we intend to use this framework to tune the parameters of additional probabilistic device types, such as tunnel diodes and other types of MTJs, as well as investigate generating samples from other probability distributions beyond exponential and gamma. Additionally, we plan to continue investigating other approaches for material and device parameter search, such as particle swarm optimization, to further understand what the tradeoffs are for each optimization approach. Finally, we also plan to extend to circuit and system design while codesigning the devices for various application types, providing a full-stack optimization paradigm.

\section{Method}
For the RL optimization strategy, we are leveraging OpenAI's RL framework, Stable-Baselines3~\cite{raffin2021stable}. This framework allowed us to build a custom environment that integrates the device models to allow agents to explore the parameter search space to optimize against KL divergence and energy consumption. The framework also provides built-in RL algorithms which simplify the training process by easily allowing the switching of training algorithms. As discussed before, our RL approach utilizes the PPO algorithm, which was developed by OpenAI and integrated into their Stable-Baselines3 framework. The framework also supports visualization tools which allowed us to monitor the training progress and test against different trained models.

For the EA optimization strategy, we utilized the Library for Evolutionary Algorithms in Python (LEAP)~\cite{coletti2020library}. LEAP is a general-purpose framework for evolutionary algorithms that employs a pipelining feature for search and optimization algorithms with useful distribution and visualization features. They provide multi-objective optimization algorithms, such as NSGA-II~\cite{deb2002fast}, which we utilized in this work. The framework allows for easy modification of population, training generations, mutation, and crossover parameters---key aspects of evolutionary algorithms---which allowed us to test various training setups.

\section{Supplementary Information}\label{Supplementary_Info}

\subsection{RL Design} \label{RL_info}

As previously stated, for each of the device models solving for the RNG application, a similar RL workflow was constructed to train an agent to optimize the materials and parameters of the two device models to best exhibit the desired distribution. There are various components that comprise this RL setup.

The first component in defining the environments for the device models is establishing the action space the agents will be operating under. The action space determines the possible actions the agent can take for a given state. For our application, the actions are the valid values of the various device parameters for the respective device model, as shown in Table~\ref{tab:table1}. A continuous action space is used for the agent, therefore, it can pick any value within the given range for each parameter, while modifying all parameters simultaneously in every step of the environment. Since each parameter has a specific valid range, the parameters are normalized to exhibit values between 0 and 1. This normalization step helps to stabilize the learning process by avoiding parameter values in drastically different ranges. Once an agent determines its action (an array of normalized parameter values), each normalized parameter value undergoes the inverse operation to extract the true parameter value in its respective range. These ``unnormalized" parameter values are then fed into the device model for evaluation.

The second component involves defining the observation space which relays the observations for a given state, so the agent can take more informed actions. For our scenario, the observations comprised the current device parameters for the state, the Config\_Score (a numerical value used to rank the configuration) for the current parameters, and the best configuration score encountered so far. The idea is that the agent knows its current standing and the goal to beat.

Finally, we can define the reward schema, which is critical to the success of the RL implementation. For our setup, the reward schema must reflect our application requirements: mitigating energy consumption of the device models, closely fitting the desired distributions, and producing viable configurations that are physically realizable and that exhibit valid behavior. 

Let us first discuss a measure of distribution fit. There are many methods to compare one distribution to another, including curve fitting methods. One metric of statistical interest is the KL divergence, which is a measure of relative entropy, roughly describing how far away a given distribution $P$ is from another distribution $Q$. It can be thought of as to how surprised one would be using $P$ as a distribution when one meant to use $Q$. Though not a true measure of distance, it is directly relatable to hypothesis testing on distributions~\cite{yang2018robust}. Furthermore, it has previously been used to empirically compare sampled data to distributions~\cite{aimone2021assessing}. Here, it is particularly useful over other statistical tests, e.g., discrete sets like $\chi^2$, because it allows us to make sure our test data $P$ never deviates too strongly from our intended distribution $Q$.

In its discrete form, the KL divergence is
\begin{equation}\label{eq:kl_div}
    \KL\left(P\,\middle\|\,Q\right) = \sum_{i=1}^{N} P(i)\log{\frac{P(i)}{Q(i)}.}
\end{equation}
Here, $P$ and $Q$ are distributions over $N$ potential outcomes or categories. The observed and expected frequency of the $i^{\text{th}}$ category is given by $P(i)$ and $Q(i)$, respectively.

We are now equipped to define the Config\_Score, which provides a quantitative measurement of how well a particular configuration is performing and makes it easier to compare and rank configurations. To calculate the Config\_Score, the respective device model is sampled $2,500$ times to calculate the average energy consumption and to compare the sampled and target distributions via KL divergence, as in
\begin{equation}\label{eq:config_score}
    \begin{split}
        \text{Config\_Score} & =  \omega_1 (\text{Avg. Energy/Coinflip})
         + \omega_2 \left ( \KL \right ) 
    \end{split}.
\end{equation}
Note that the weight values ($\omega_1, \omega_2$) are used to prioritize certain behaviors: either incentivizing energy efficiency or more closely matching the desired distribution. This directly ties into how the parameter values are modified by the agent in order to best maximize its rewards. Based on the weighting, certain parameter combinations that are responsible for the desired behavior will be more strongly motivated.

There are also checks in place to determine if the device configuration is valid. The checks sample the respective device models and monitor whether the magnetic free layer is stimulated appropriately to exhibit RNG behavior. This includes sampling the device configuration to characterize the probability of sampling a 1 as a function of the current and magnetic field being applied. This curve is then tested to see if it exhibits a sigmoidal ``S-curve'' trend. We can now define the reward function as shown in Eq.~\ref{eq:reward_func}.
\begin{equation}\label{eq:reward_func}
    \text{Reward}=
    \begin{cases}
        -1 & \text{if invalid configuration} \\
        1 & \text{if Config\_score $<$ Best\_Config\_Score} \\
        0 & \text{otherwise}
    \end{cases}
\end{equation}
If the checks fail, the agent is penalized with a negative reward. Since the goal is to minimize the energy consumption and deviations from the target distributions, the agent acquires a positive reward when it produces a configuration where the Config\_Score minimizes the smallest Config\_Score discovered thus far. Otherwise, a reward of zero is given. This simplified reward scheme was adopted rather than having the configuration score acting as the reward because it allows the agent to be rewarded when it performs a desired action---finding a better configuration. This should help prevent the agent from falling in a local optimum to accrue rewards and encourage the agent to find better solutions. For both devices, the agent was trained for 6,000 timesteps and then tested on 150 episodes with each episode consisting of 60 timesteps.

\subsection{Gamma Distribution Selection}

As previously mentioned, the gamma distribution appears in a variety of fields. We selected our parameters based on a hypothetical one-dimensional particle tracking experiment. Inference for diffusion processes is a subject area useful for real-life particle tracking. Such inference techniques are discussed in great detail in~\cite{fuchs2013inference}. For our hypothetical particle tracking experiment, imagine one is allowed to track a physical particle diffusing in some fluid over time. One might think of a microscope experiment, perhaps in the classical sense of tracking pollen particles in water. For simplicity, we limit ourselves to a particle diffusing in one dimension. We want to estimate the coefficient of viscous drag of the tracked particle in the fluid.

We let $X_t$ denote the position of our particle over time and $V_t$ denote the velocity of the particle over time. Adopting stochastic differential equation (SDE) notation, the way position varies over time is described by
\begin{equation}
    \rmd X_t = V_t\rmd t.
    \label{eq:part_pos}
\end{equation}
Assuming our particle is freely diffusing in fluid, there are no active forces acting on the particle. Rather, the particle is subject to thermal diffusion and viscous drag. We will further assume that the particle's mass is not changing over time, which may happen if the particle were eroding due to dissolution or friction. In the Langevin sense, the forces acting on the particle can be represented as follows:
\begin{equation}
    m\rmd V_t = -\alpha V_t\rmd t+ \sqrt{2k_BT\alpha}\rmd W_t,
    \label{eq:force_sde}
\end{equation}
where $m$ represents the mass of the particle, $\alpha$ the coefficient of viscous drag, $k_B$ Boltzmann's constant, and $T$ the absolute temperature. The process $W_t$ is a standard Brownian motion with zero mean and unit variance. Equations~\ref{eq:part_pos} and~\ref{eq:force_sde} describe the dynamics of our particle.

When the mass of the particle is sufficiently small, Eq.~\ref{eq:force_sde} increases in variance. Therefore, we take the over-damped limit\footnote{See~\cite{pavliotis2008multiscale} for averaging and homogenization techniques for SDEs.} and obtain the single equation for position:
\begin{equation}
    \rmd X_t = \sqrt{\frac{2k_BT}{\alpha}}\rmd W_t.
    \label{eq:particle_sde}
\end{equation}
For sufficiently small time intervals $\Delta t$, the quantity $\zeta_i = X_{t_i+\Delta t}-X_{t_i}$ is normally distributed with mean zero and variance $2k_BT\Delta t/\alpha$. Moreover, for sufficiently small $\Delta t$, these $\zeta_i$ represent independent increments.\footnote{See~\cite{fuchs2013inference} for additional details on independent increments and inference with diffusion observations.}

We will assume that for this experiment, $T$ is known. The only unknown in Eq.~\ref{eq:particle_sde} is then $\alpha$. Assume that we observe our particle at fixed intervals of $\Delta t$ and obtain the collection of $n$ displacement measurements $\left\{\zeta_i\right\}$. Therefore, since we know these are independent and identically distributed as previously described, we can write down the likelihood density of seeing those measurements given the value of the parameter $\alpha$:
\begin{align}
    \begin{split}
        \mathcal{L}\left(\left\{\zeta_i\right\}\,\middle|\,\alpha\right) & = \prod_{i=1}^n\sqrt{\frac{\alpha}{4\pi k_BT\Delta t}}\exp\left(-\frac{\alpha}{4k_BT\Delta t}\zeta_i^2\right),\\
        &=\left(\frac{\alpha}{4\pi k_BT\Delta t}\right)^{\frac{n}{2}}\exp\left(-\frac{\alpha}{4k_BT\Delta t}\sum_{i=1}^n\zeta_i^2\right).
    \end{split}
    \label{eq:likelihood}
\end{align}

This likelihood density can be used in a Bayesian estimation fashion to estimate $\alpha$ given the collection of observations $\left\{\zeta_i\right\}$. To do so, we must sample the posterior distribution $\pi\left(\alpha\,\middle|\,\left\{\zeta_i\right\}\right)\stackrel{c}{=}\mathcal{L}\left(\left\{\zeta_i\right\}\,\middle|\,\alpha\right)\pi_0\left(\alpha\right)$, where $\pi_0\left(\alpha\right)$ is the prior distribution on $\alpha$ and $\stackrel{c}{=}$ denotes equality up to a normalizing constant. We select a Jeffreys prior for $\alpha$. That is, we select $\pi_0\left(\alpha\right)=1/\alpha$. This prior allows us to remain mostly uniformed on what $\alpha$ should be while simultaneously specifying that it must be positive. We will now manipulate the posterior distribution, pulling all multiplicative terms not involving $\alpha$ into the normalizing constant.
\begin{align}
    \begin{split}
        \pi\left(\alpha\,\middle|\,\left\{\zeta_i\right\}\right) &\stackrel{c}{=} \left(\frac{\alpha}{4\pi k_BT\Delta t}\right)^{\frac{n}{2}}\exp\left(-\frac{\alpha}{4k_BT\Delta t}\sum_{i=1}^n\zeta_i^2\right)\pi_0\left(\alpha\right),\\
        &\stackrel{c}{=}\alpha^{\frac{n}{2}-1}\exp\left(-\left(\frac{1}{4k_BT\Delta t}\sum_{i=1}^n\zeta_i^2\right)\alpha\right).
    \end{split}
    \label{eq:posterior_alpha}
\end{align}
In this form, with multiplicative constants absorbed, we see that the posterior distribution for $\alpha$ is a gamma density. Therefore, we conclude $\alpha$ is gamma distributed with shape $n/2$ and rate $-(1/4k_BT\Delta t)\sum_{i=1}^n\zeta_i^2$.

Hence, if we were executing a particle tracking experiment and wanted to estimate the coefficient of viscous drag $\alpha$ using our measurements, we could sample or utilize the statistics from a gamma distribution with these parameters to estimate the mean and uncertainty of $\alpha$.

\begin{figure}[htp]
\centering
\includegraphics[width=4in]{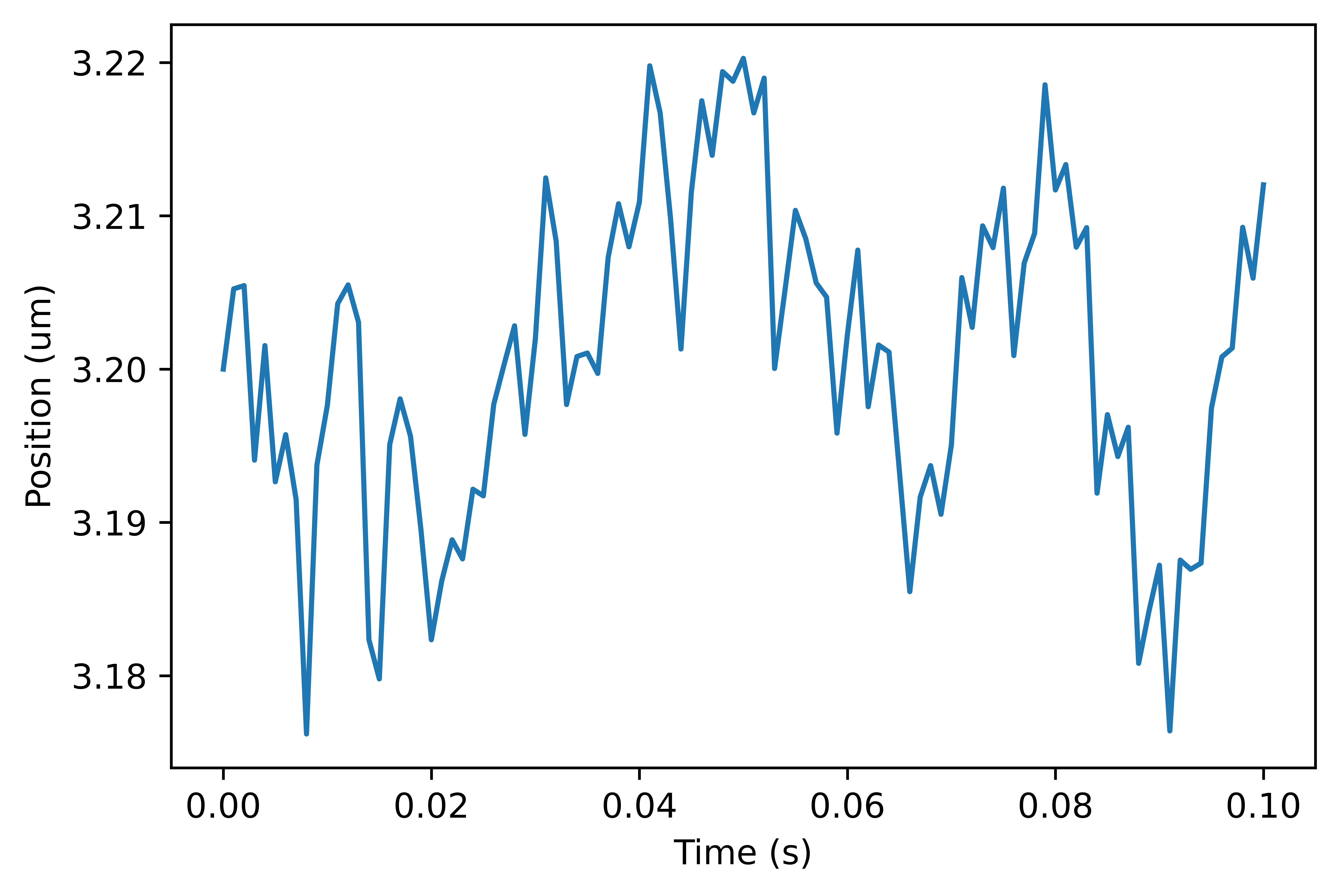}
\caption{Simulated particle trajectory (Eq.~\ref{eq:particle_sde}) for $100$ timesteps with $\Delta t = 0.001$ s, $X_0 = 3.2$ pN, $\alpha= 0.16$ pN s$/\mu$m, and $k_BT = 0.0041$ pN $\mu$m. From Eq.~\ref{eq:posterior_alpha}, this data implies $\alpha$ is a gamma distributed random variable with shape $50.00$ and rate approximately equal to $311.44$ $\mu$m$/($pN s$)$.}
\label{fig:particle_trajectory}
\end{figure}

To generate a gamma distribution to sample from, we simulated Eq.~\ref{eq:particle_sde} for 100 timesteps starting from $X_0 =3.2$ pN with $\alpha = 0.16$ pN s$/\mu$m, $k_BT=0.0041$ pN $\mu$m, and $\Delta t=0.001$ s. This trajectory is plotted in Fig.~\ref{fig:particle_trajectory}. We calculated the intervals from this simulated particle tracking experiment and used them to calculate the shape and rate of the gamma distribution. The shape parameter is $50.00$ (unitless) and the rate parameter is approximately $311.44$ $\mu$m$/($pN s$)$.\footnote{Note that the theoretical mean of the gamma distribution is the ratio of shape to rate. Here, that would imply the mean of $\alpha$ is approximately $0.16054$ pN s$/\mu$m. This is not exactly the true value of $\alpha=0.16$ pN s$/\mu$m. However, this is still a pretty good approximation given the small sample size of a single observation for 100 timesteps.}

We use these parameters to define the gamma distribution sampled in the main text. While these are tied to this synthetic particle tracking application, they are sufficiently distinct from previously explored distributions. Namely, this distribution provides a tightly clustered asymmetric mass away from the origin---in direct contrast to the previously explored exponential distribution.

\subsection{Alternate Device Selection}

\begin{figure}[htp]
\centering
\includegraphics[width=0.8\textwidth]{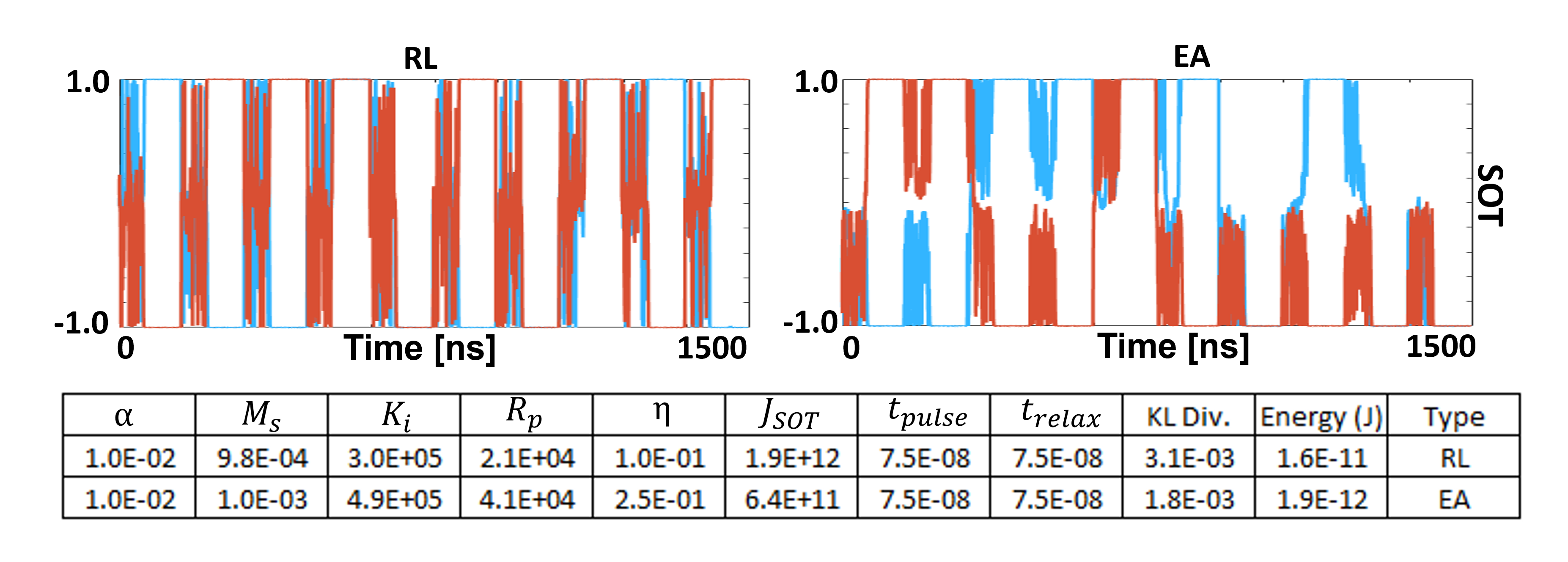}
\caption{Sample bitstreams for devices `pushed out' of SOT operation and into stochastic operation by the optimization strategies.}
\label{fig:nonSOT}
\end{figure}

While most of the SOT devices selected by the EA and RL approaches exhibit expected behavior, it is noted that both optimization strategies do at times drive the device out of SOT operation and into a stochastic operation state. As an example, consider the two discovered configurations shown in Fig.~\ref{fig:nonSOT}. While the effect of the applied SOT current to overcome the PMA of the devices and move the magnetization away from $\pm z$ can clearly be seen, it does not settle neatly in-plane. Instead, the magnetization moves about quickly (possibly with some bias, as in the shown EA bitstream) until the SOT current stops, at which point it relaxes into $\pm z$. Even though these devices operate in a different way, both bitstreams still achieve low KL divergence, indicating that they are both still valid TRNG devices. In fact, these particular configurations have slightly lower KL divergence than the best-performing SOT devices, though consume slightly more energy. This highlights the capability of the optimization strategies to find additional ways of operating devices outside the intended scheme. If these alternate devices are not desired to be found, additional checks can be incorporated into the optimization strategies to discourage or even fully filter out these cases.

\subsection{Device Variations and Probability Bias}

\begin{figure}[htp]
\centering
\includegraphics[width=\textwidth]{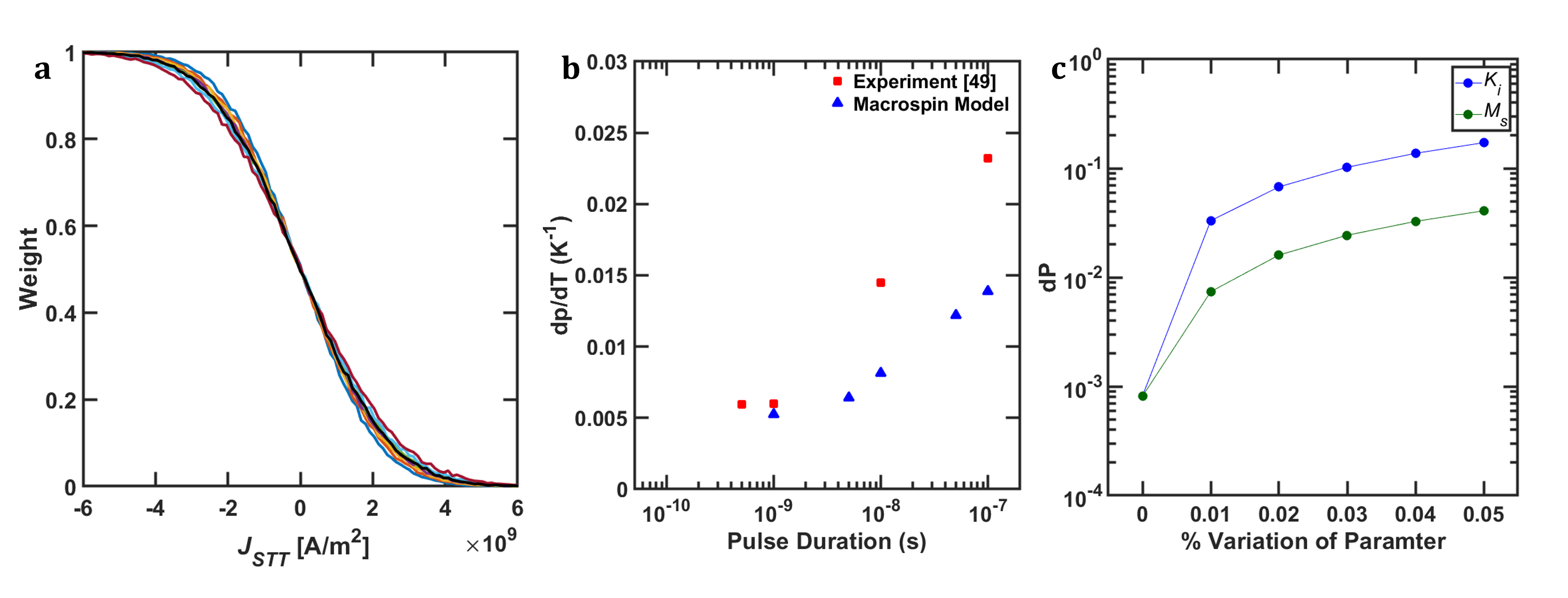}
\caption{ (a) When inducing small variations in the device parameters (such as anisotropy, saturation magnetization, etc.) the device S-curves warp, leading to alternate biasing with applied STT currents (for both SOT and STT device types). (b) The predicted change in probability from 50\% ($dP$) of an unbiased STT device operating under temperature fluctuations ($dT$) according to the macrospin model is compared against measurement data for SMART-MTJ devices found in~\cite{rehm2024temperature} While the model, in general, underestimates the amount of probability change, the overall trends agree, and there is particularly good agreement when pulse duration is low. (c) Predicted magnitude of the change in probability from 50\% of an unbiased SMART-MTJ device due to small variations in the device saturation magnetization $M_s$ and anisotropy $K_i$.}
\label{fig:validation}
\end{figure}

In this section, we briefly discuss the effects of variation of the device parameters on device probability bias, including comparison to measurement results. As stated in the main text, a single device with known parameters can be characterized by an S-curve which describes the relationship between the magnitude of the applied biasing (SOT-devices) or switching (STT-devices) current $J_{STT}$ and the resulting probability bias. When device parameters vary, the S-curve relationship changes for each varied device. Figure~\ref{fig:validation}a shows this change in S-curve for a handful of randomly generated SOT devices with device parameters (such as $M_s$ or $K_i$) within 5\% of each other. As the AI-guided codesign framework tests many hypothetical devices, it is important to generate these S-curves for each novel device in order to accurately capture the bias for generating gamma-distributed random numbers.

Another way to capture the change in biasing behavior of the devices due to variations in the device parameters is shown in Figures~\ref{fig:validation}b-c. In Figure~\ref{fig:validation}b we model the change in probability $dP$ due to a change in temperature $dT$ for SMART-MTJ devices from~\cite{rehm2024temperature}, which are stochastic STT-MTJ devices, and compare the measurement data to the results of the macrospin model. While temperature fluctuations were not considered in this article (all devices were simulated at room temperature), the available measurement data for comparison and existence of analytical bounds~\cite{rehm2024temperature} allow for macrospin model verification. Specifically, the 50\% bias point is first found for the devices at different pulse durations. Then for each pulse duration, a temperature change of $10 K$ is induced, changing the bias of the device. While the macropsin model in general underestimates the change in probability bias due to a change in temperature, the overall trends match, and importantly both the model and the measurement data approach the low-pulse duration limit~\cite{rehm2024temperature}.

A similar analysis occurs in Figure~\ref{fig:validation}c, where we now hold pulse duration constant but induce increasing variations in device parameters $M_s$ or $K_i$. The figure shows that variations in either parameter change the biasing behavior of the devices, with variations in $K_i$ more strongly affecting the bias than $M_s$. Additional studies varying other parameters can be performed as well, as in~\cite{morshed2023reduced}.\\

\noindent Please refer to our source code for implementation details:\\
\url{https://github.com/utinclab/Stochastic_MTJ_Model}

\section*{Acknowledgements}
We acknowledge support from the DOE Office of Science (ASCR / BES) Microelectronics Co-Design project COINFLIPS. We thank Lindsey Aimone for copyediting assistance.

This paper describes objective technical results and analysis. Any subjective views or opinions that might be expressed in the paper do not necessarily represent the views of the U.S. Department of Energy or the United States Government. Sandia National Laboratories is a multimission laboratory managed and operated by National Technology \& Engineering Solutions of Sandia, LLC, a wholly owned subsidiary of Honeywell International Inc., for the U.S. Department of Energy’s National Nuclear Security Administration under contract DE-NA0003525. SAND Number: SAND2024-15031O.

\end{document}